\documentclass[12pt,a4paper]{article}

\usepackage{fixltx2e}

\usepackage[font=small]{caption}
\usepackage[text={450pt,650pt},headheight={14.5pt},centering]{geometry}

\usepackage{lmodern}

\usepackage{graphicx}
\usepackage[subrefformat=parens,labelformat=parens]{subfig}
\usepackage{tikz}
\usetikzlibrary{trees}
\usetikzlibrary{decorations.pathmorphing}
\usetikzlibrary{decorations.markings}
\usepackage{booktabs}
\usepackage{array}
\usepackage{multirow}
\usepackage{dsfont}
\usepackage[sort,compress]{cite}
\usepackage{collref}

\usepackage{amsmath,amssymb}
\usepackage{mathtools}
\usepackage{slashed}
\usepackage{braket}

\usepackage{hyperref}

\usepackage{xspace}

\numberwithin{equation}{section}

\makeatletter
 \let\old@startsection=\@startsection
 \let\oldl@section=\l@section
 \renewcommand{\@startsection}[6]{\old@startsection{#1}{#2}{#3}{#4}{#5}{#6\mathversion{bold}}}
 \renewcommand{\l@section}[2]{\oldl@section{\mathversion{bold}#1}{#2}}
\makeatother

\renewcommand{\leq}{\leqslant}
\renewcommand{\geq}{\geqslant}

\def\XXint#1#2#3{{\setbox0=\hbox{$#1{#2#3}{\int}$}
    \vcenter{\hbox{$#2#3$}}\kern-.5\wd0}}

\newcommand{\AdS}{\text{AdS}}
\newcommand{\CFT}{\text{CFT}}
\newcommand{\Sphere}{\text{S}}
\newcommand{\Torus}{\text{T}}

\newcommand{\Smat}{\mathcal{S}}

\newcommand{\alg}[1]{\mathfrak{#1}}

\newcommand{\algSL}{\alg{sl}}

\newcommand{\algSU}{\alg{su}}

\newcommand{\algU}{\alg{u}}

\newcommand{\algPSU}{\alg{psu}}

\newcommand{\gen}[1]{\mathfrak{#1}}

\newcommand{\Reals}{\mathbb{R}}

\newcommand{\order}{\mathcal{O}}

\newcommand{\ie}{\textit{i.e.}\xspace}
\newcommand{\eg}{\textit{e.g.}\xspace}
\newcommand{\cf}{\textit{cf.}\xspace}

\newcommand{\CDD}{\scriptscriptstyle\text{CDD}}

\newcommand{\am}{\text{am}}
\newcommand{\sn}{\text{sn}}
\newcommand{\cn}{\text{cn}}
\newcommand{\dn}{\text{dn}}
\newcommand{\K}{\text{K}}

\newcommand{\Fup}{F^{\rotatebox[origin=c]{360}{$\scriptstyle\curvearrowleft$}}}
\newcommand{\Fdw}{F^{\rotatebox[origin=c]{180}{$\scriptstyle\curvearrowleft$}}}

\renewcommand{\Re}{\mathop{\operatorname{Re}}}
\renewcommand{\Im}{\mathop{\operatorname{Im}}}

\DeclareMathOperator{\sign}{sign}

\newcommand{\mat}[1]{\mathbb{#1}}
\newcommand{\matId}{\mathds{1}}

\newcommand{\prodsigma}{\sigma^+}
\newcommand{\ratiosigma}{\sigma^-}
\newcommand{\sumtheta}{\theta^+}
\newcommand{\difftheta}{\theta^-}
\newcommand{\BES}{\text{BES}}
\newcommand{\BLMMT}{\text{BLMMT}}
\newcommand{\AFS}{\text{AFS}}
\newcommand{\HL}{\text{HL}}
\newcommand{\newq}{\mathcal{Q}}
\newcommand{\curvearrowurl}{\curvearrowleft}
\newcommand{\curvearrowdlr}{\rotatebox[origin=c]{180}{$\curvearrowleft$}}
\newcommand{\inturl}{\;{\int \negthickspace \negthickspace \negthickspace\negthickspace \negthinspace \curvearrowurl}\mbox{ }\,} 
\newcommand{\intdlr}{\;{\int \negthickspace \negthickspace \negthickspace \negthinspace \curvearrowdlr}\mbox{ }\,} 
\newcommand{\ointc}{\;{\int \negthickspace \negthickspace \negthickspace \negthinspace \circlearrowleft}\mbox{ }\,}

\begin{document}
\tikzset{
particle/.style={draw=blue,line width=1, postaction={decorate},
    decoration={markings,mark=at position .5 with {\arrow[draw=blue]{>}}}},
antiparticle/.style={draw=blue,line width=1, postaction={decorate},
    decoration={markings,mark=at position .5 with {\arrow[draw=blue]{<}}}},
particlecross/.style={draw=red,line width=1,dashed, postaction={decorate},
    decoration={markings,mark=at position .5 with {\arrow[draw=red]{>}}}},
antiparticlecross/.style={draw=red,line width=1,dashed, postaction={decorate},
    decoration={markings,mark=at position .5 with {\arrow[draw=red]{<}}}},
nopart/.style={draw=none}
 }

\begin{flushright}\footnotesize\ttfamily
DMUS-MP-13/14\\
ITP-UU-13/14\\
SPIN-13/10
\end{flushright}
\vspace{4em}

\begin{center}
\textbf{\Large\mathversion{bold} Dressing phases of $\AdS_3$/$\CFT_2$}

\vspace{2em}

\textrm{\large Riccardo Borsato${}^1$, Olof Ohlsson Sax${}^1$, Alessandro Sfondrini${}^1$,\\ Bogdan Stefa\'nski, jr.${}^2$ and Alessandro Torrielli${}^3$} 

\vspace{3em}

\begingroup\itshape
1. Institute for Theoretical Physics and Spinoza Institute, Utrecht University, Leuvenlaan 4, 3584 CE Utrecht, The Netherlands

2. Centre for Mathematical Science, City University London, Northampton Square, EC1V 0HB, London, UK

3. Department of Mathematics, University of Surrey, Guildford, GU2 7XH, UK\par\endgroup

\vspace{1em}

\texttt{R.Borsato@uu.nl, O.E.OlssonSax@uu.nl, A.Sfondrini@uu.nl, Bogdan.Stefanski.1@city.ac.uk, a.torrielli@surrey.ac.uk}


\end{center}

\vspace{6em}

\begin{abstract}\noindent
We determine the all-loop dressing phases of the $\AdS_3/\CFT_2$ integrable system related to type IIB string theory on $\AdS_3\times \Sphere^3\times \Torus^4$ by solving the recently found crossing relations and studying their singularity structure.
The two resulting phases present a novel structure with respect to the ones appearing in $\AdS_5/\CFT_4$ and $\AdS_4/\CFT_3$. In the strongly-coupled regime, their leading order reduces to the universal Arutyunov-Frolov-Staudacher phase as expected. We also compute their sub-leading order and compare it with recent one-loop perturbative results, and comment on their weak-coupling expansion.
\end{abstract}

\newpage

\tableofcontents

\section{Introduction}

The gauge/string correspondence~\cite{Maldacena:1997re,Gubser:1998bc,Witten:1998qj} is a remarkable relation between quantum gauge and gravity theories. In the planar limit~\cite{'tHooft:1973jz} of certain dual pairs, the correspondence can be understood in terms of an integrable system with the 't~Hooft coupling constant $\lambda$ entering as a free parameter.\footnote{For a comprehensive review and list of references see~\cite{Beisert:2010jr}. A detailed exposition of integrable string theory on $\AdS_5$ can be found in~\cite{Arutyunov:2009ga}.} At small values of $\lambda$ the integrable system reduces to an integrable spin-chain with local interactions~\cite{Minahan:2002ve,Beisert:2003jj}. At large values of $\lambda$, in the thermodynamic limit, the integrable system is described by a set of integral equations known as the finite-gap equations~\cite{Kazakov:2004qf,Beisert:2005bm}; these integral equations can be obtained using the classical (Lax) integrability of the string theory equations of motion~\cite{Bena:2003wd}. 

Integrable systems' S-matrices satisfy the Yang-Baxter equation, which allows for an arbitrary scalar factor in its solution. Fixing this factor requires imposing additional constraints, the most powerful of these being crossing symmetry. In the context of   the $\AdS_5/\CFT_4$ integrable system, crossing symmetry constraints were first identified in~\cite{Janik:2006dc}. The solution of these constraints~\cite{Beisert:2006ib,Beisert:2006ez,Kostov:2008ax,Dorey:2007xn,Arutyunov:2009kf,Volin:2009uv}, the so-called dressing phase, conventionally written as $\sigma\equiv e^{i\theta}$, is a key ingredient in matching the strong and weak coupling limits of the dualilty~\cite{Arutyunov:2004vx}. 

In the $\AdS_5/\CFT_4$ and $\AdS_4/\CFT_3$ integrable systems a non-perturbative dressing phase was found by Beisert, Eden and Staudacher in~\cite{Beisert:2006ez}; we will denote it by $\sigma^\BES$.  At small coupling the dressing phase is trivial at the leading two orders: $\sigma^\BES=1+{\cal O}(\lambda^3)$, while for large $\lambda$ it appears already at the leading order. It is conventional to refer to the first two orders in the strong coupling expansion of a dressing phase as the Arutyunov-Frolov-Staudacher (AFS)~\cite{Arutyunov:2004vx} and Hern\'andez-L\'opez (HL)~\cite{Hernandez:2006tk,Freyhult:2006vr} orders, which appear at (${\cal O}(\lambda^{1/2})$ and ${\cal O}(\lambda^0)$), respectively. Expanding $\sigma^\BES$ at strong coupling, one finds the leading AFS phase~\cite{Arutyunov:2004vx} which is expected to be universal for many integrable string theory backgrounds. The next-to-leading term gives the HL phase. This was first found through a one-loop sigma-model computation~\cite{Beisert:2005cw,Hernandez:2006tk,Freyhult:2006vr} and can also be obtained through a semi-classical quantisation of the finite gap equations~\cite{Gromov:2007cd}. This latter derivation shows explicitly that the HL phase is the same for all states in a given background. Finally, since the dressing phase is the same in $\AdS_5/\CFT_4$ and $\AdS_4/\CFT_3$, one may ask whether $\sigma^\BES$ is a universal dressing phase of $\AdS/\CFT$ integrable systems.

The $\AdS_3/\CFT_2$ correspondence~\cite{Brown:1986nw} for theories with 16 supercharges has recently been investigated using integrability methods. There are in fact two distinct classes of $\AdS_3$ backgrounds with this amount of supersymmetry: $\AdS_3\times \Sphere^3\times \Torus^4$ and $\AdS_3\times \Sphere^3\times \Sphere^3\times \Sphere^1$; these were studied following Maldacena's seminal paper, see for example~\cite{Maldacena:1998bw,Seiberg:1999xz,Larsen:1999uk,
Gauntlett:1998kc,Elitzur:1998mm,deBoer:1999rh,Gukov:2004ym}. 

The integrablity approach to the $\AdS_3/\CFT_2$ correspondence was initiated in~\cite{Babichenko:2009dk}. Building on the actions~\cite{Metsaev:1998it,Stefanski:2007dp,Arutyunov:2008if,Stefanski:2008ik} (see also~\cite{Pesando:1998wm, Park:1998un, Metsaev:2000mv, Gomis:2008jt}), string theories, in a certain kappa-gauge, on such $\AdS_3$ backgrounds were shown to be classically integrable~\cite{Babichenko:2009dk}.\footnote{Classical integrability has also been investigated independent of any kappa-gauge fixing~\cite{Sorokin:2010wn,Sundin:2012gc}.} The finite gap equations~\cite{Babichenko:2009dk} and conjectured all-loop Bethe Ansaetze  were written down in~\cite{Babichenko:2009dk,OhlssonSax:2011ms}. However, this procedure keeps track only of the excitations which remain massive in the BMN limit~\cite{Berenstein:2002jq}. Fully incorporating the massless modes remains an open issue; for recent progress on this, see~\cite{Sax:2012jv}.

 In~\cite{Borsato:2012ud,Borsato:2012ss} an S-matrix and Bethe Ansatz for the $\AdS_3\times \Sphere^3\times \Sphere^3\times \Sphere^1$ integrable system was written down.  Expanding on this, in~\cite{Borsato:2013qpa} the exact integrable S-matrix and associated Bethe Ansatz of the integrable system associated to Type IIB string theory on $\AdS_3\times \Sphere^3\times \Torus^4$ with R-R flux was constructed.\footnote{For other work in this direction see~\cite{David:2008yk,David:2010yg,Ahn:2012hw}.} A number of recent papers have performed important perturbative calculations for string theory on $\AdS_3$ backgrounds~\cite{Rughoonauth:2012qd,Sundin:2012gc,
Abbott:2012dd,Beccaria:2012kb,Beccaria:2012pm,Sundin:2013ypa,Bianchi:2013nra,Engelund:2013fja}.  Further, the integrability of a family of  $\AdS_3\times \Sphere^3\times \Torus^4$ theories with both R-R and NS-NS flux has been investigated~\cite{Cagnazzo:2012se,Hoare:2013pma,Hoare:2013ida}. Integrability appears also to play an important role in $\AdS_3$ black-hole solutions~\cite{David:2012aq}.

The integrable system related to Type IIB string theory on $\AdS_3\times \Sphere^3\times \Torus^4$ with R-R flux has two dressing phases, as can be expected on fairly general grounds~\cite{Hoare:2011fj,Borsato:2013qpa}.\footnote{This is quite natural since in $\AdS_3/\CFT_2$ the left- and right- movers are independent.} The phases satisfy two  crossing relations~\cite{Borsato:2013qpa} (see equation~(\ref{eq:crossing12}) below).\footnote{By inspection,  there is no straightforward linear combination of the two phases whose crossing relation would reduce to the $\AdS_5$ crossing relations~\cite{Janik:2006dc}.} One can check that these crossing relations imply that the two dressing phases behave differently under double-crossing, and therefore the two phases must be distinct. In this paper we find the non-perturbative dressing phases of the $\AdS_3/\CFT_2$ S-matrix by solving the crossing relations. We identify the bound states of the system and show that the full S-matrix including these dressing phases has the right pole structure to account for such bound states. 

We perform a strong coupling expansion of our dressing phases and find that at the AFS-order both dressing phases are the same as the AFS-phase of  $\AdS_5/\CFT_4$ and  $\AdS_4/\CFT_3$, confirming its universality. This is in agreement with an explicit strong-coupling regime calculation done in~\cite{Rughoonauth:2012qd,Sundin:2013ypa,Beccaria:2012kb}. At the HL-order, our phases are different from one another, confirming that the system does have two distinct dressing phases. Only the sum of these two phases is the same as the HL-phase of $\AdS_5/\CFT_4$ and $\AdS_4/\CFT_3$. This shows that, unlike the AFS-phase, the HL-order phase is {\em not} universal. We find that at the HL-order our dressing phases are almost, though not quite, the same as the ones obtained in~\cite{Beccaria:2012kb}. 
We discuss the possible origins of this discrepancy and how it may be resolved.\footnote{We would like to thank Matteo Beccaria,  Fedor Levkovich-Maslyuk, Guido Macorini,  and Arkady Tseytlin for discussions about this point.} In the near-flat space limit our dressing phases agree with the results of~\cite{Sundin:2013ypa}. We also give a weak-coupling expansion of our dressing phases.

This paper is organised as follows. 
In section~\ref{sec:torus-crossing} we review the crossing relations derived in~\cite{Borsato:2013qpa} and their interpretation in terms of the rapidity torus. 
In section~\ref{sec:crossing-solution} we solve the crossing relations non-perturbatively. 
In section~\ref{sec:poles}, we analyse the BPS bound state spectrum and the corresponding singularities of the S-matrix.
In section~\ref{sec:expansions} we perform a strong-coupling expansion of the two phases, compare with the results of~\cite{Beccaria:2012kb,Sundin:2013ypa} and give the weak coupling expansion of the phases. Some technical results are relegated to the appendices.

\bigskip

\noindent
{\bf Note added:}
Shortly after this paper, another work appeared~\cite{Abbott:2013mpa}, where a semiclassical derivation of dressing phases was performed for the $\AdS_3\times \Sphere^3\times \Sphere^3\times \Sphere^1$ and $\AdS_3\times \Sphere^3\times \Torus^4$ backgrounds, taking into account some issues of anti-symmetrisation, cutoffs and surface terms. The results for $\AdS_3\times \Sphere^3\times \Torus^4$ agree with the Hern\'andez-L\'opez order of our proposal, and the ones for $\AdS_3\times \Sphere^3\times \Sphere^3\times \Sphere^1$ are half of that, irrespectively of the masses of the excitations. This is compatible with the crossing (and in particular double-crossing) equations of~\cite{Borsato:2012ud}, which suggests that the~$\AdS_3\times \Sphere^3\times \Sphere^3\times \Sphere^1$ phases may be found in terms of the ones presented here, even at all-loop. We plan to return to this issue in the near future.

\section{Rapidity torus and crossing equations}\label{sec:torus-crossing}
In this section we will consider the crossing equations of~\cite{Borsato:2013qpa} on the rapidity torus where the $\AdS_3\times \Sphere^3\times \Torus^4$ dispersion relation is uniformized. 

\subsection{Uniformizing the dispersion relation}
The all-loop dispersion relation for massive excitations on $\AdS_3\times \Sphere^3\times \Torus^4$ reads~\cite{Borsato:2013qpa}\footnote{%
  We express the dispersion relation in terms of the coupling constant $h(\lambda)$, which is related to the world-sheet coupling $\lambda$ by
  \begin{equation}
    h(\lambda) = \frac{\sqrt{\lambda}}{4\pi} + \order(1/\sqrt{\lambda}) .
  \end{equation}
  In~\cite{Beccaria:2012kb} it was shown that there is no $\order(1)$ term in the strong coupling expansion of $h(\lambda)$.
}%
\begin{equation}
E(p)=\sqrt{1+16h^2\sin^2\frac{p}{2}}\,.
\end{equation}
In analogy with the relativistic case, it is convenient to introduce a rapidity variable $z$ which uniformizes the dispersion relations. In the present setting, given the similarity with the $\AdS_5\times \Sphere^5$ dispersion relations, the rapidity will also live on a complex torus~\cite{Janik:2006dc}. Following the conventions of~\cite{Arutyunov:2009ga}, let us define
\begin{equation}
p(z)=2\,\am z\,,\qquad \sin\frac{p(z)}{2}=\sn(z,\kappa)\,,\qquad E(z)=\dn(z,\kappa)\,,
\end{equation}
in terms of Jacobi's elliptic functions, where the elliptic modulus is $\kappa=-16h^2$. This defines a torus with a real period $2\omega_1$ and an imaginary period $2\omega_2$ that depend on $h$ through
\begin{equation}
\omega_1=2\,\K(\kappa)\,,\qquad
\omega_2=2i\,\K(1-\kappa)-2\,\K(\kappa)\,,
\end{equation}
where $\K$ is the complete elliptic integral of the first kind. The Zhukovski variables~$x^\pm(z)$ are meromorphic functions on the torus
\begin{equation}
x^\pm(z)=\frac{1}{4h}\left(\frac{\cn(z,\kappa)}{\sn(z,\kappa)}\pm i\right)\left(1+\dn(z,\kappa)\right)\,.
\end{equation}
They satisfy the shortening condition~\cite{Borsato:2013qpa}
\begin{equation}
\label{eq:short}
\left(x^+(z)+\frac{1}{x^+(z)}\right)-\left(x^-(z)+\frac{1}{x^-(z)}\right)=\frac{i}{h}\,.
\end{equation}
One can check that the Zhukovski variables and the dispersion relations are also $\omega_1$-periodic, so that we can always restrict to $|\Re (z)|\leq \omega_1/2$, corresponding to $-\pi < p \leq \pi$. In this parameterization the real $z$-axis lies in the physical region, since it corresponds to real momentum and positive energy. 

The crossing transformation corresponds to changing the sign of momentum and energy. In terms of $x^\pm(z)$ this is achieved by sending $x^\pm\to1/x^\pm$, which amounts to a shift by half of the imaginary period of the rapidity torus, $z\to z\pm \omega_2$. For a meromorphic function on the torus shifting up or down makes no difference, but the S-matrix that we are interested in will not be meromorphic on the product of two such tori. In fact, due to the presence of the dressing factors, we expect it to have and infinite number of cuts there.

\begin{figure}
  \centering
  \subfloat[Torus with $|x^\pm|=1$ curves]{%
    \label{fig:torusabs}
    \includegraphics[width=50mm]{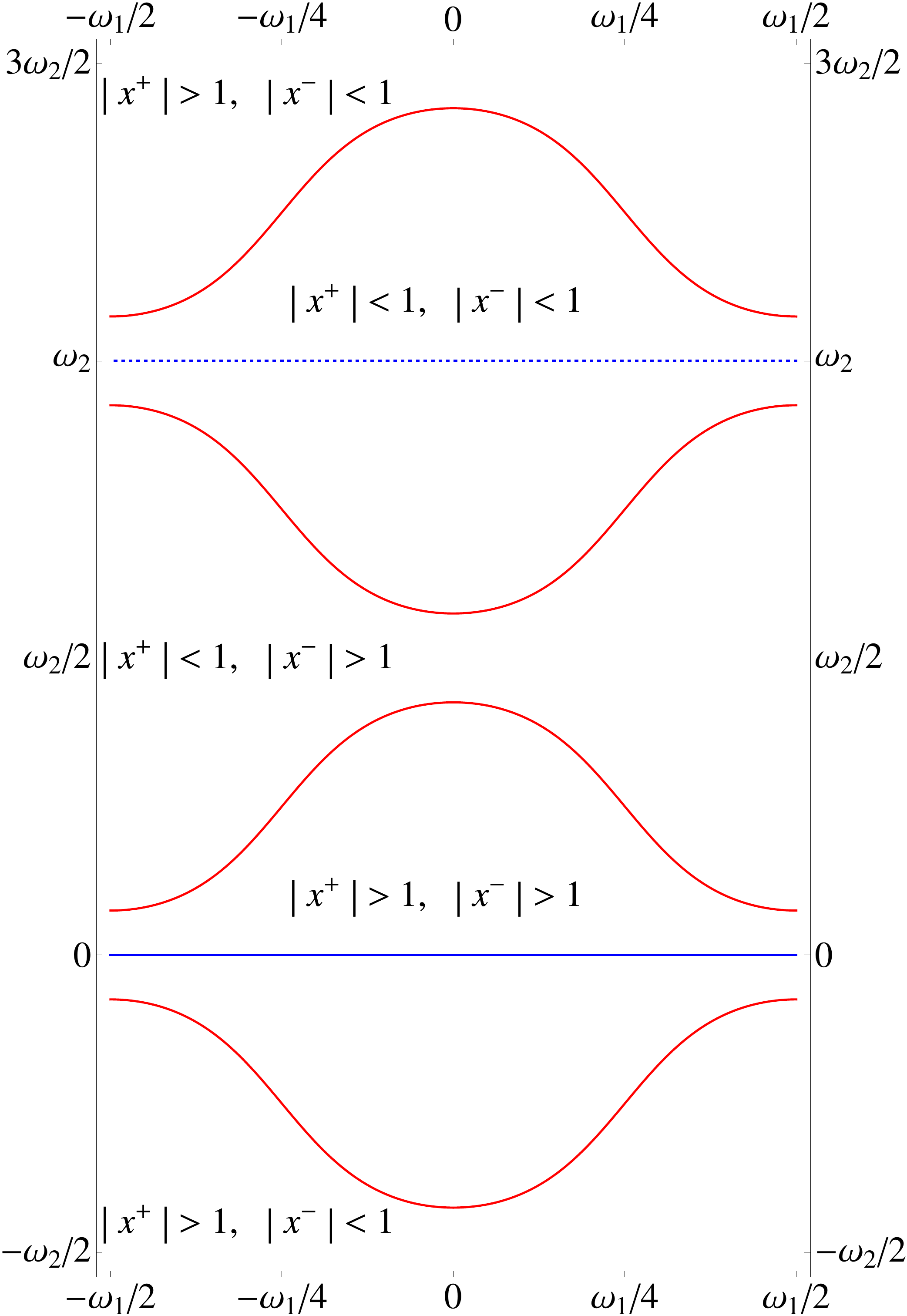}}
  \subfloat[Torus with $\Im(x^\pm)=0$ curves]{%
    \label{fig:torusim}
    \includegraphics[width=50mm]{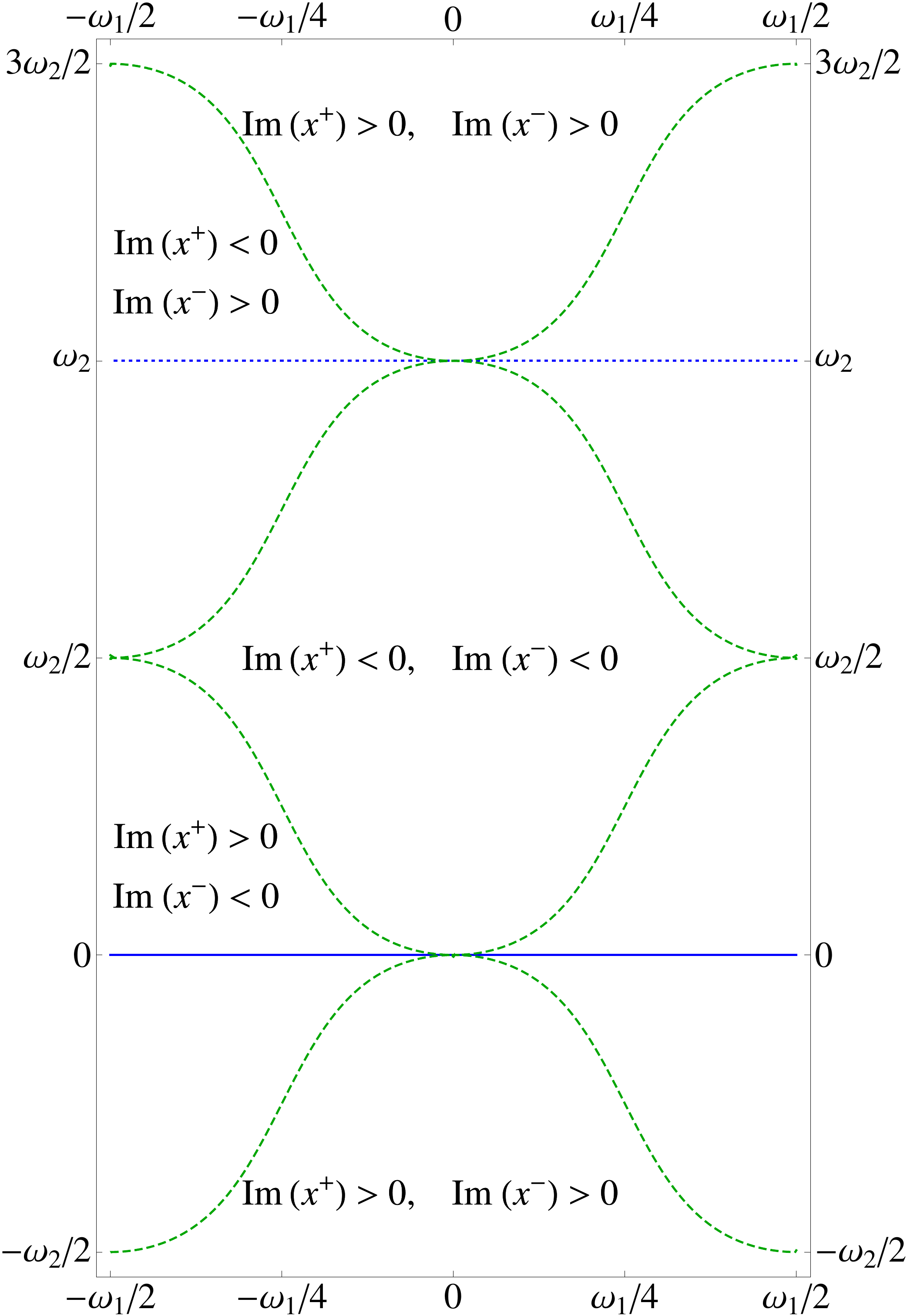}}
  \subfloat[Torus with both curves]{%
    \label{fig:torusabsim}
    \includegraphics[width=50mm]{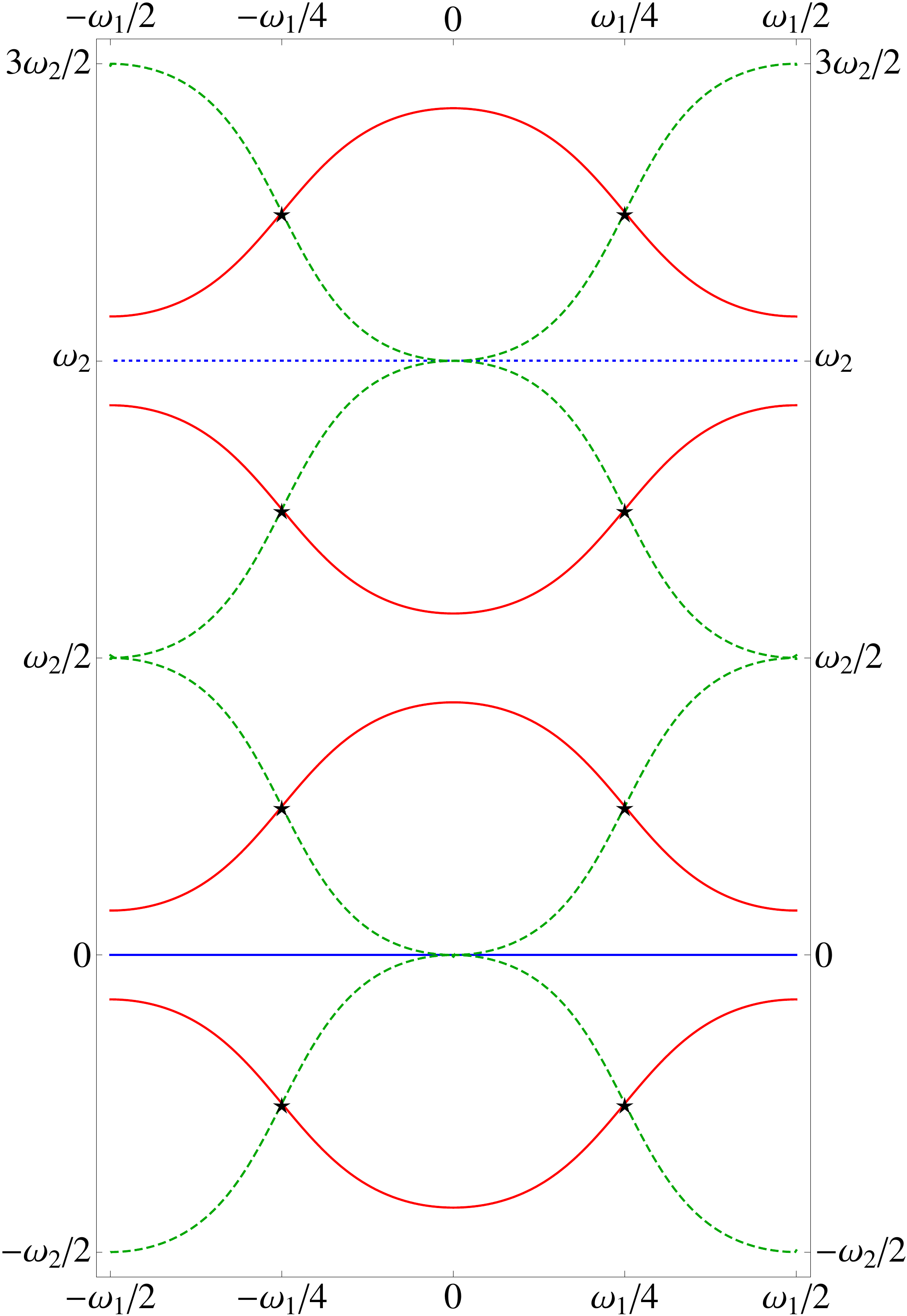}}
  \caption{%
    The rapidity torus with several significant curves. The solid blue line is the real $z$-axis (physical region),  the dashed blue line is the $z=\omega_2$ axis (``crossed'' region). In the leftmost figure the torus is divided in four regions by  $|x^\pm|=1$ and  in the central figure it is divided by $\Im(x^\pm)=0$. The rightmost picture depicts both sets of curves, which intersect in eight points with real part $\pm\omega_1/4$.
  }%
  \label{fig:torus}
\end{figure}

Based on the experience with the $\AdS_5\times \Sphere^5$ case it is convenient to identify distinct regions on the $z$-torus. The curves $|x^\pm(z)|=1$ divide the torus into four non-intersecting regions, depicted in figure~\subref*{fig:torusabs}, and so do the curves $\Im(x^\pm)=0$, see figure~\subref*{fig:torusim}. These two sets of curves intersect in eight points that lie on $\Re(z)=\pm\omega_1/4$, see figure~\subref*{fig:torusabsim}.

\subsection{The crossing equations}
In~\cite{Borsato:2013qpa} the all-loop S-matrix for $\AdS_3\times \Sphere^3\times \Torus^4$ strings was proposed. It contained two undetermined dressing factors $\sigma(p_1,p_2)$ and $\widetilde{\sigma}(p_1,p_2)$. Unitarity and physical unitarity constrain them to be of the form
\begin{equation}
\label{eq:twoads3phases}
\sigma(p_1,p_2)=e^{i\,\theta(p_1,p_2)},\qquad
\widetilde{\sigma}(p_1,p_2)=e^{i\,\tilde{\theta}(p_1,p_2)},
\end{equation} 
where $\theta(p_1,p_2)$ and $\widetilde{\theta}(p_1,p_2)$ are antisymmetric real analytic functions for real $p_1,\,p_2$.

It was also shown that crossing invariance requires these factors to obey a set of crossing equations,
\begin{equation}
\begin{aligned}
\sigma(\bar{p}_1,p_2)^2\,\widetilde{\sigma}(p_1,p_2)^2= g(p_1,p_2)\,,&\qquad&
\sigma(p_1,p_2)^2\,\widetilde{\sigma}(\bar{p}_1,p_2)^2= \widetilde{g}(p_1,p_2)\,,\\
\sigma(p_1,\bar{p}_2)^2\,\widetilde{\sigma}(p_1,p_2)^2= \frac{1}{\widetilde{g}(\bar{p}_2,p_1)}\,,&\qquad&
\sigma(p_1,p_2)^2\,\widetilde{\sigma}(p_1,\bar{p}_2)^2= \frac{1}{g(\bar{p}_2,p_1)}\,,
\end{aligned}
\label{eq:crossing12}
\end{equation}
where the bar indicates crossing and
\begin{equation}
  \begin{aligned}
    g(p_1,p_2) &= \left(\frac{x_2^-}{x_2^+}\right)^2\frac{\left(1-\frac{1}{x_1^+x_2^+}\right)\left(1-\frac{1}{x_1^-x_2^-}\right)}{\left(1-\frac{1}{x_1^+x_2^-}\right)^2}\frac{x_1^- - x_2^+}{x_1^+ - x_2^-}\,,\\
    \widetilde{g}(p_1,p_2) &= \left(\frac{x_2^-}{x_2^+}\right)^2\frac{\left(x_1^- - x_2^+\right)^2}{\left(x_1^+ - x_2^+\right)\left(x_1^- - x_2^-\right)}\frac{1-\frac{1}{x_1^-x_2^+}}{1-\frac{1}{x_1^+x_2^-}}\,.
\end{aligned}
\label{eq:crossingF}
\end{equation}
Antisymmetry requires that in~\eqref{eq:crossing12} the shift by $\omega_2$ is done in opposite directions in the first and the second variables of the dressing factors; this leaves us with two distinct choices. Fixing the direction of the shift amounts to choosing a path for analytic continuation of the dressing phase from the physical to the crossed region. As discussed in appendix~\ref{sec:an-cont}, compatibility with the perturbative results requires
\begin{equation}
\bar{p}_1\equiv p(z_1+\omega_2)\,,\qquad
\bar{p}_2\equiv p(z_2-\omega_2)\,,
\end{equation}
This is the same convention as in the case of $\AdS_5\times \Sphere^5$~\cite{Arutyunov:2009ga}.

By iterating the crossing transformation twice we find that the dressing factors are not $2\omega_2$-periodic
\begin{equation}
\begin{aligned}
\frac{\sigma(z_1+2\omega_2,z_2)^2}{\sigma(z_1,z_2)^2}=\frac{g(z_1+\omega_2,z_2)}{\widetilde{g}(z_1,z_2)} &= \left(\frac{x_1^+ - x_2^+}{x_1^+ - x_2^-}\frac{x_1^- - x_2^-}{x_1^- - x_2^+}\right)^2\,,\\
\qquad
\frac{\widetilde{\sigma}(z_1+2\omega_2,z_2)^2}{\widetilde{\sigma}(z_1,z_2)^2}=\frac{\widetilde{g}(z_1+\omega_2,z_2)}{g(z_1,z_2)} &= \left(\frac{1-\frac{1}{x_1^+x_2^-}}{1-\frac{1}{x_1^+x_2^+}}\frac{1-\frac{1}{x_1^-x_2^+}}{1-\frac{1}{x_1^-x_2^-}}\right)^2\,.
\end{aligned}
\end{equation}
This confirms our expectation that the dressing factors have cuts on the rapidity torus.

We are interested in solving the crossing equations
\begin{equation}
\label{eq:crossing1}
\sigma(z_1+\omega_2,z_2)^2\,\widetilde{\sigma}(z_1,z_2)^2= g(z_1,z_2)\,,\qquad
\sigma(z_1,z_2)^2\,\widetilde{\sigma}(z_1+\omega_2,z_2)^2= \widetilde{g}(z_1,z_2)\,.
\end{equation}
Our result will be manifestly antisymmetric, so that the crossing equations in the second variable will automatically follow.

\begin{figure}
  \centering
  \includegraphics[width=70mm]{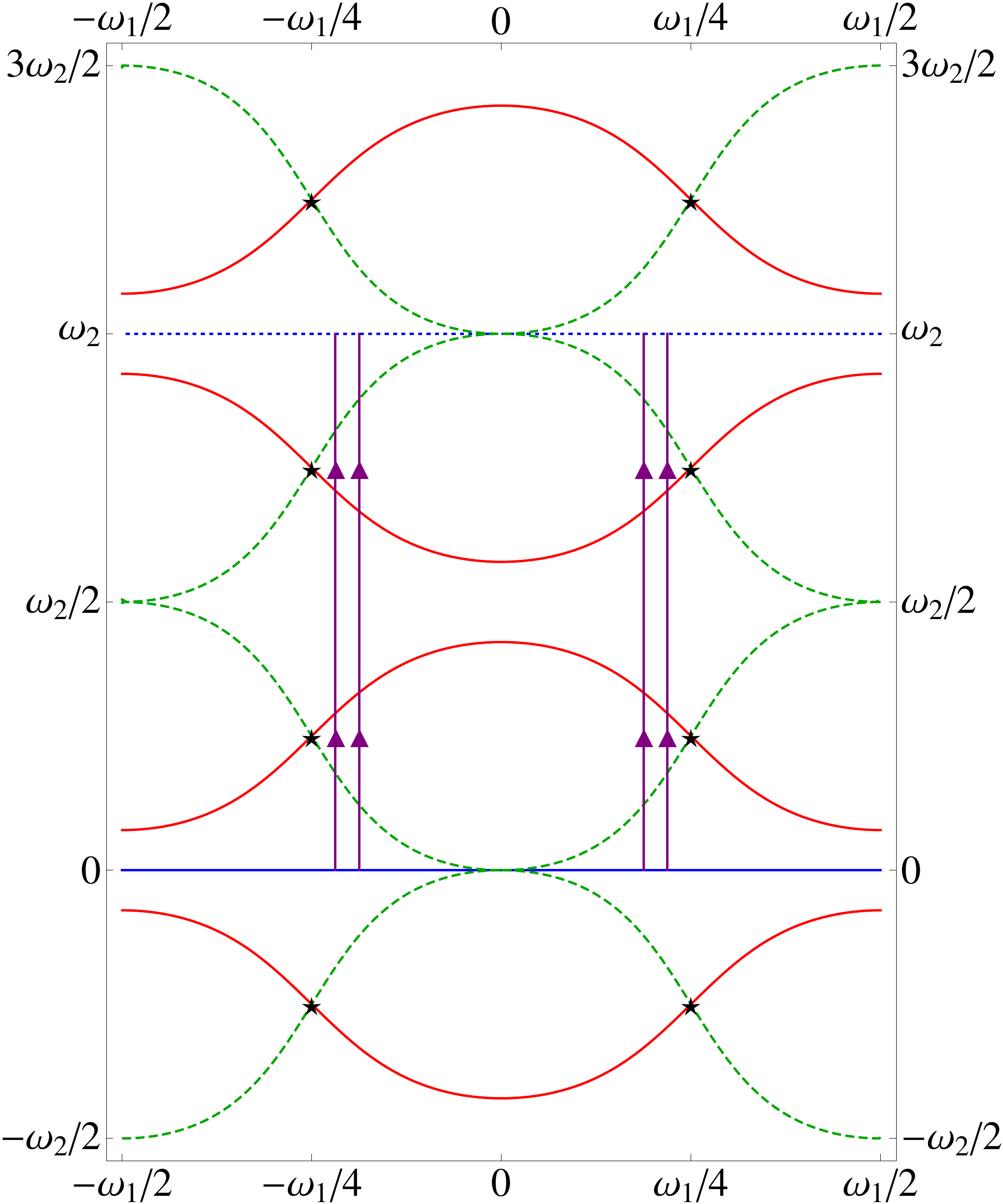}
  \caption{%
    The paths used for analytic continuation from $z$ to $z+\omega_2$ (in purple) are vertical segments that lie close to the boundary of $|\Re(\gamma)|<\omega_1/4$. They cross the red lines $|x^\pm|=1$ when $\Im(x^\pm)<0$.
  }%
  \label{fig:toruspaths}
\end{figure}

We have to give a prescription for performing the continuation from $z$ to $z+\omega_2$ on the torus. In order to construct the dressing factors $\sigma$ and $\widetilde{\sigma}$ we will exploit some properties of the Beisert-Eden-Staudacher (BES) phase~\cite{Beisert:2006ez},
which compels us to choose a path compatible with the crossing transformation for that phase. This has been discussed in detail in~\cite{Arutyunov:2009kf}, and we will follow the procedure outlined there.
Figure~\ref{fig:toruspaths} depicts the paths we will use to reach the crossed region along curves $\gamma(z)$ that go from $z$ to $z+\omega_2$ with constant $\Re (\gamma)$, which lie close to the boundaries of the region $|\Re(\gamma)|<\omega_1/4$ and crossing the lines $|x^\pm|=1$ in the region $\Im (x^\pm) <0$.\footnote{%
  This is a subset of the paths used in the case of $\AdS_5$, see section 4 in~\cite{Arutyunov:2009kf}.%
} %

\section{Solutions of the crossing equations}\label{sec:crossing-solution}

In order to solve~\eqref{eq:crossing1} we will consider the crossing equations for the sum and the difference of the two phases  $\theta(p_1,p_2)$ and~$\widetilde{\theta}(p_1,p_2)$.
Let us denote the product  and the ratio of the dressing factors by
\begin{equation}
\prodsigma(p_1,p_2)\equiv \sigma(p_1,p_2)\, \widetilde{\sigma}(p_1,p_2)\,,\qquad\ratiosigma(p_1,p_2)\equiv \frac{\sigma(p_1,p_2)}{\widetilde{\sigma}(p_1,p_2)}\,,
\end{equation}
and corresponding phases by~$\sumtheta(p_1,p_2)$ and~$\difftheta(p_1,p_2)$. It is also useful to rewrite each phase as~\cite{Arutyunov:2006iu}
\begin{equation}
\label{eq:thchi}
\theta(p_1,p_2) = \chi(x_1^+,x_2^+) +\chi(x_1^-,x_2^-) -\chi(x_1^+,x_2^-) -\chi(x_1^-,x_2^+)\,,
\end{equation}
where $\chi$ is an antisymmetric function, with similar expressions  for~$\widetilde{\theta}(p_1,p_2),\, \sumtheta(p_1,p_2)$ and~$\difftheta(p_1,p_2)$.

\subsection{The BES and HL phases}
In what follows we will use some properties of the BES~\cite{Beisert:2006ez} and HL phases~\cite{Hernandez:2006tk}, which we briefly recall here. The $\AdS_5$ S-matrix contains a single dressing phase, which satisfies the crossing equation~\cite{Janik:2006dc} 
\begin{equation}\label{eq:cr-BES}
\begin{gathered}
\sigma^\BES(z_1,z_2) \sigma^\BES(z_1+\omega_2,z_2) = h(x_1^\pm,x_2^\pm), \\
h(x_1^\pm,x_2^\pm) \equiv\frac{x_2^-}{x_2^+}\frac{x_1^- - x_2^+}{x_1^- - x_2^-} \frac{1-\frac{1}{x_1^+x_2^+}}{1-\frac{1}{x_1^+x_2^-}}.
\end{gathered}
\end{equation}
The solution is given by the BES phase~\cite{Beisert:2006ez}. A particularly useful representation of this phase in the physical region was given by Dorey, Hofman and Maldacena (DHM)~\cite{Dorey:2007xn}
\begin{equation}
\chi^\BES(x,y)= i \ointc \frac{dw}{2 \pi i} \ointc \frac{dw'}{2 \pi i} \, \frac{1}{x-w}\frac{1}{y-w'} \log{\frac{\Gamma[1+i h(w+1/w-w'-1/w')]}{\Gamma[1-i h(w+1/w-w'-1/w')]}}\,. 
\label{eq:besdhmrep}
\end{equation}
The leading term in the strong coupling limit ($h\rightarrow \infty$) of this phase\footnote{%
  To find the asymptotic expansion of the BES phase at strong coupling one can expand the integrand using that $i\log\frac{\Gamma(1+ix)}{\Gamma(1-ix)}=-x\log\frac{x^2}{e^2}-\frac{\pi}{2} \sign(\Re x)-2\sum_{n=0}^\infty \frac{\zeta(-2n-1)}{2n+1}\frac{(-1)^n}{x^{2n+1}}$ for $\Re x\neq0$. This expression corrects some typos in the expansion given in~\cite{Vieira:2010kb}.%
} %
is given by the Arutyunov-Frolov-Staudacher (AFS) phase~\cite{Arutyunov:2004vx}
\begin{equation}
\label{eq:AFS-xpxm}
\sigma^\AFS(x_1,x_2) = \left(  \frac{1-\frac{1}{x_1^-x_2^+}}{1-\frac{1}{x_1^+x_2^-}} \right) \left(  \frac{1-\frac{1}{x_1^+x_2^-}}{1-\frac{1}{x_1^+x_2^+}}  \frac{1-\frac{1}{x_1^-x_2^+}}{1-\frac{1}{x_1^-x_2^-}} \right)^{i h (x_1+1/x_1-x_2-1/x_2)}\,,
\end{equation}
while the next-to-leading order correction is the Hern\'andez-L\'opez (HL) phase~\cite{Hernandez:2006tk}, 
\begin{equation}\label{eq:DHM-HL}
\chi^\HL(x,y)= \frac{\pi}{2} \ointc \frac{dw}{2 \pi i} \ointc \frac{dw'}{2 \pi i} \, \frac{1}{x-w}\frac{1}{y-w'} \, \text{sign}(w'+1/w'-w-1/w)\,.
\end{equation}
For later convenience, let us perform one of the two integrals in (\ref{eq:DHM-HL}) and obtain the representation
\begin{equation}\label{eq:intu-intd-HL}
\chi^\HL(x,y)=\left( \inturl - \intdlr \right)\frac{dw}{4\pi} \frac{1}{x-w} \left(\log{(y-w)}-\log{\left(y-1/w\right)}\right),
\end{equation}
where the two integrals are performed in the upper and lower unit semi-circle respectively, counterclockwise in both cases.

The HL phase solves the ``odd'' part of the $\AdS_5$ crossing equation~\cite{Beisert:2006ib}
\begin{equation}\label{eq:cr-HL}
\sigma^\HL(z_1,z_2) \sigma^\HL(z_1+\omega_2,z_2) = \sqrt{ \frac{h_{12}}{h_{\bar{1}2}} }=\sqrt{h_{12}\,(h_{12})^*}\,,
\end{equation}
where
\begin{equation}
h_{12}\,(h_{12})^* =\frac{\ell^\HL(x_1^+,x_2^-)\,\ell^\HL(x_1^-,x_2^+)}{\ell^\HL(x_1^+,x_2^+)\,\ell^\HL(x_1^-,x_2^-)}\,, \qquad \ell^\HL(x,y)\equiv\frac{x-y}{1-xy}\,,
\end{equation}
where complex conjugation amounts to sending $x_k^\pm \rightarrow x_k^\mp$.\footnote{One can check that in order for~\eqref{eq:intu-intd-HL} to solve~\eqref{eq:cr-HL} it is necessary to choose the path of analytic continuation as in figure~\ref{fig:toruspaths}. To do this one can mimic the arguments presented in appendix~\ref{app:identities}.}

\subsection{Solution for the sum of the phases}\label{sec:sol-diff}
Taking the product of the two crossing equations~(\ref{eq:crossing1}), we find an equation for $\prodsigma$
\begin{equation}
{\prodsigma(z_1,z_2)}^2 \, {\prodsigma(z_1+\omega_2,z_2)}^2 = g_{12} \, \widetilde{g}_{12}\,.
\label{eq:crosssum}
\end{equation}
We observe that the r.h.s of this equation can be written in terms of the function $h_{12}$ appearing on the r.h.s of the $\AdS_5$ crossing equation~(\ref{eq:cr-BES}) 
\begin{equation}
g_{12} \, \widetilde{g}_{12}=\frac{(h_{12})^3}{(h_{12})^*}\,,
\end{equation}
where the shortening condition~(\ref{eq:short}) is used. The above relation allows us to solve the crossing equation~(\ref{eq:crosssum}) using parts of the $\AdS_5$ dressing phase
\begin{equation}
\label{eq:sumsolution}
\prodsigma_{12}=\frac{(\sigma^{\BES}_{12})^2}{\sigma^{\HL}_{12}}\,, \qquad\text{\ie}\qquad \sumtheta_{12}=2\theta^{\BES}_{12}-\theta^{\HL}_{12}
\,.
\end{equation}
To show that $\prodsigma_{12}$ defined in this way satisfies equation~(\ref{eq:crosssum}) one need only use equations~\eqref{eq:cr-BES} and~\eqref{eq:cr-HL}. It is convenient to express $\prodsigma_{12}$ in terms of a DHM-like double-integral representation, by defining $\chi^+(x,y)$ as
\begin{equation}
\label{eq:chi+}
\begin{aligned}
\chi^+(x,y)=&\,2\,\chi^{\BES}(x,y)-\chi^{\HL}(x,y)\\
 =&\ointc \frac{dw}{2 \pi i} \ointc \frac{dw'}{2 \pi i} \, \frac{1}{x-w}\frac{1}{y-w'} \Bigg( 2i \log{\frac{\Gamma[1+i h(w+1/w-w'-1/w')]}{\Gamma[1-i h(w+1/w-w'-1/w')]}} \\
&\hspace{6cm}- \frac{\pi}{2} \text{sign}(w'+1/w'-w-1/w) \Bigg)\,,
\end{aligned}
\end{equation}
in the physical region. Notice that the above expression is exact to all orders in the coupling $h$.

Let us comment on the strong-coupling behaviour of the above expression for $\prodsigma$. In the next subsection we will show that the difference of the two phases has no term at the AFS order. Taking this into account, it follows that both $\sigma$ and $\tilde{\sigma}$ reduce to $\sigma^{\AFS}$ at leading order, as expected on general grounds and confirmed by the explicit calculations in~\cite{Beccaria:2012kb}.
Moreover, this result confirms that $\theta+\tilde{\theta}$ is equal to $\theta^\HL$ -- the $\AdS_5$ HL term, as predicted by~\cite{Beccaria:2012kb}.

\subsection{Solution for the difference of the phases}
Taking the ratio of the two crossing equations~(\ref{eq:crossing1}), we get
\begin{equation}\label{eq:cr-ratio}
\frac{{\ratiosigma(z_1,z_2)}^2}{{\ratiosigma(z_1+\omega_2,z_2)}^2}=\frac{\widetilde{g}_{12}}{g_{12}}\,,
\end{equation}
where
\begin{equation}\label{eq:cr-ratio-expl}
\frac{\widetilde{g}_{12}}{g_{12}}=\frac{\ell^-(x_1^+,x_2^-)\ell^-(x_1^-,x_2^+)}{\ell^-(x_1^+,x_2^+)\ell^-(x_1^-,x_2^-)}, \qquad \ell^-(x,y)\equiv(x-y)\left(1-\frac{1}{xy}\right).
\end{equation}
Notice that this equation involves the ratio rather than the product of the dressing factor with its analytic continuation.
As we show in appendix~\ref{app:identities-proof}, defining $\chi^-(x,y)$ in the physical region ($|x|,|y|>1$) by the integral
\begin{equation}\label{eq:chi-}
\begin{aligned}
\chi^-(x,y) &=\ointc \, \frac{dw}{8\pi} \frac{1}{x-w} \log{\left[ (y-w)\left(1-\frac{1}{yw}\right)\right]} \, \text{sign}((w-1/w)/i) \ - x \leftrightarrow y \\
&=\left( \inturl - \intdlr \right)\frac{dw}{8\pi} \frac{1}{x-w} 
\log{\left[ (y-w)\left(1-\frac{1}{yw}\right)\right]} \ - x \leftrightarrow y\,,
\end{aligned}
\end{equation}
solves the crossing equation~\eqref{eq:cr-ratio}. By construction, $\chi^-$ is antisymmetric. As mentioned in the previous subsection, in the strong coupling expansion $\chi^-$ is zero at leading (AFS) order; at the next-to-leading (HL) order it has a non-zero term, which we discuss in detail in section~\ref{sec:expansions}. Finally, we note that the integrand of $\chi^-$ has a trivial expansion in the coupling $h$, in contrast to the solution of the crossing equation~(\ref{eq:crosssum}) which is solved by an integrand with a non-trivial series expansion in $h$.  This is because equation~\eqref{eq:cr-ratio} is  ``odd'' in the sense of~\cite{Beisert:2006ib}.

The all loop expressions for $\chi$ and $\tilde{\chi}$ are then given by
\begin{equation}
\label{eq:solution}
\begin{aligned}
  \chi(x,y) &= \chi^{\text{BES}}(x,y)+\frac{1}{2}\left(-\chi^{\text{HL}}(x,y)+\chi^{-}(x,y)\right) \,, \\
  \widetilde{\chi}(x,y) &= \chi^{\text{BES}}(x,y)+\frac{1}{2}\left(-\chi^{\text{HL}}(x,y)-\chi^{-}(x,y)\right) \,.
\end{aligned}
\end{equation}
These solutions are expressed in terms of the non-perturbative BES phase plus terms at the HL order. These latter contributions to $\chi$ and $\tilde{\chi}$ are independent of $h$. As such, they can be added to the DHM representation of the BES  phase without affecting the $h$-resummation.

\section{Single poles and bound states}\label{sec:poles}

There is a close connection between simple poles in the physical region of the S-matrix and the bound states of the model. In this section we will first give a brief overview of the expected bound state spectrum, and then discuss the corresponding simple poles.

\subsection{Short representations}

The ground state is preserved by the algebra $\algPSU(1|1)^4 \times \algU(1)^4$, where the four $\algU(1)$ factors represent central charges~\cite{Borsato:2012ud,Borsato:2013qpa}. The excitations transform in short representations of this algebra, satisfying a shortening condition relating the four central charges by~\cite{Borsato:2012ud}
\begin{equation}\label{eq:su11-shortening}
  \gen{H}^2 - 4\gen{P}\gen{P}^\dag = \gen{M}^2 \,.
\end{equation}
Bound states preserving some supersymmetry also transform in a short representation of the symmetry algebra. Let us consider the two-particle state\footnote{%
  There is no loss in generality in doing so, since the shortening condition~\eqref{eq:su11-shortening} is expressed in terms of central charges.%
} %
$\ket{\Phi^{+\dot{+}}_p \Phi^{+\dot{+}}_q}$ containing two left-moving bosons.\footnote{%
  Following the notation of~\cite{Borsato:2013qpa} the left-moving bosonic excitations are $\Phi^{+\dot{+}}$ and $\Phi^{-\dot{-}}$, and correspond to excitations on $\AdS_3$ and $\Sphere^3$, respectively. Similarly, the right-moving bosons are $\bar{\Phi}^{+\dot{+}}$ and $\bar{\Phi}^{-\dot{-}}$. The distinction between the ``left-moving'' and ``right-moving'' (or L and R) excitations comes from the fact that $\Phi^{-\dot{-}}$ carries positive angular momentum on $\AdS_3$, while the angular momentum of $\bar{\Phi}^{-\dot{-}}$ is negative. Hence, these excitations can be thought of as left- and right-movers in the dual $\CFT_2$.%
} %
For generic values of the momenta $p$ and $q$ the tensor product of two fundamental left-moving representations is an irreducible long representation. However, at special points the tensor product becomes reducible. In particular, we find that the shortening condition~\eqref{eq:su11-shortening} is satisfied for $x_p^+ = x_q^-$ and $x_p^- = x_q^+$. Only at these points it is possible to construct short sub-representations. Therefore any pole in the S-matrix corresponding to a \emph{supersymmetric} bound state will have to satisfy one of these conditions.

An interesting feature of the $\algPSU(1|1)^4 \times \algU(1)^4$ algebra is that all short irreducible representations are two-dimensional while all long irreducible representations have dimension four. A two-particle bound state will therefore transform in a representation which has the same form as the fundamental representation, differing only in the values of the central charges. This should be contrasted with the centrally extended $\algPSU(2|2)$ algebra appearing in $\AdS_5 \times \Sphere^5$~\cite{Beisert:2004ry,Beisert:2005tm,Arutyunov:2006ak}, where the fundamental representation has dimension four while the $M$-particle bound state has dimension $4M$~\cite{Chen:2006gp}.

At the points where the tensor product becomes reducible some of the elements of the S-matrix become zero or develop poles. In order to fully understand the behaviour of the S-matrix at these points we will need to take the dressing phase into account. This will be further analysed in section~\ref{sec:S-matrix-poles}. Here we will instead consider the matrix structure of the S-matrix. Since some of the entries in the matrix vanish, the corresponding bound state representation is formed from the states on which the S-matrix acts non-trivially. For the point $x_p^+ = x_q^-$ we find that the state $\ket{\Phi^{+\dot{+}}_p \Phi^{+\dot{+}}_q}$ belongs to the short representation, and we will therefore refer to it as a $\algSU(2)$ bound state. In the case $x_p^- = x_q^+$ the short representation includes the state $\ket{\Phi^{-\dot{-}}_p \Phi^{-\dot{-}}_q}$, and is a potential $\algSL(2)$ bound state.\footnote{%
  In $\AdS_5 \times \Sphere^5$ the physical bound states correspond to ``$\algSU(2)$ bound states''. The ``$\algSL(2)$ bound states'' appear as  bound states of the mirror theory~\cite{Arutyunov:2007tc}.%
}%

To decide which bound state belongs to the physical spectrum we need to impose additional constraints on the momenta of the fundamental excitations.
In the region $s_1 \ll s_2$ the wavefunction of a scattering state takes the general form\footnote{%
  In order to avoid confusion with the dressing phase we denote the world-sheet coordinate by $s$.%
}%
\begin{equation}
  \Psi(s_1,s_2) = e^{i(p s_1 + q s_2)} + S(p,q) e^{i(p s_2 + q s_1)} ,
\end{equation}
where the first term describes the incoming wave and the second term the outgoing wave. To find a bound state we analytically continue the wavefunction to complex values of the momenta
\begin{equation}
  p = \frac{p'}{2} + iv , \qquad
  q = \frac{p'}{2} - iv.
\end{equation}
The wavefunction then behaves as $\Psi(s_1,s_2) \sim e^{v(s_2 - s_1)} + S(p,q) e^{-v(s_2 - s_1)}$ ($s_1 \ll s_2$). For the bound state wavefunction to be normalizable the exponential multiplying $S(p,q)$ should be decaying. Hence we are interested in the solution where the momentum of the first particle has a positive imaginary part. By solving the condition~\eqref{eq:short} for $x_p^\pm$ and $x_q^\pm$ in the physical region, we find that for $x_p^+ = x_q^-$ the momentum $p$ has a positive imaginary part, while $x_p^- = x_q^+$ leads to the imaginary part being negative. We hence conclude that only the $\algSU(2)$ bound state can appear in the physical spectrum. In section~\ref{sec:S-matrix-poles} we will check that the full S-matrix, including the correct scalar factor and the dressing phase, has a corresponding pole in the physical region.

So far we have only considered bound states in the LL-sector. If we start with two right-moving excitations we again find an $\algSU(2)$ bound state at $x_p^+ = x_q^-$, since the S-matrix in~\cite{Borsato:2013qpa} is symmetric under exchange of left- and right-movers. Let us finally consider the state $\ket{\Phi^{+\dot{+}}_p \bar{\Phi}^{-\dot{-}}_q}$ consisting of one left- and one right-moving excitations. In this case the shortening condition~\eqref{eq:su11-shortening} is satisfied for $x_p^+ = 1/x_q^+$ and $x_p^- = 1/x_q^-$. Neither of these solutions lie in the physical region $|x_p^\pm|>1$, $|x_q^\pm|>1$ and hence there are no supersymmetric bound states in the LR-sector.

In summary we find that physical two-particle  $\algSU(2)$ bound states exist in the LL- and RR-sectors. The LR-sector, on the other hand, does not contain any bound states.

\subsection{Giant magnons}

At strong coupling of the  $\AdS_5/\CFT_4$ duality the fundamental scalar excitations are described by \emph{giant magnons}~\cite{Hofman:2006xt}. The simple giant magnon is a string solution living in a $\Reals \times \Sphere^3$ subspace of $\AdS_5 \times \Sphere^5$. The giant magnon solution can be extended to a solution in $\Reals \times \Sphere^3$, the \emph{dyonic} giant magnon~\cite{Chen:2006gea}, which carries angular momentum $M$ along the additional angle. This solution corresponds to a bound state of $|M|$ fundamental magnons~\cite{Dorey:2006dq}. 

Since both the fundamental giant magnon and the dyonic extension live in $\Reals \times \Sphere^3$ they can be directly embedded in $\AdS_3 \times \Sphere^3$~\cite{Abbott:2012dd}. In $\AdS_5 \times \Sphere^5$ a dyonic giant magnon with positive $\algU(1)$ charge $M$ can be continuously rotated to the corresponding magnon with negative charge $-M$. However, in the case of $\AdS_3 \times \Sphere^3$ such a rotation is not possible since the intermediate states would not sit inside $\Sphere^3$, so the two states with charges $+M$ and $-M$ are independent. In the scalar sector the left- and right-moving excitations are distinguished by the sign of the angular momentum $M$. Hence, the dyonic giant magnons with charges $+M$ and $-M$ correspond to bound states of $|M|$ left- and right-moving fundamental excitations, respectively. Setting $M=2$ we find exactly the two $\algSU(2)$ bound states expected from the representation theory considerations above.

\subsection{Simple poles of the S-matrix}
\label{sec:S-matrix-poles}

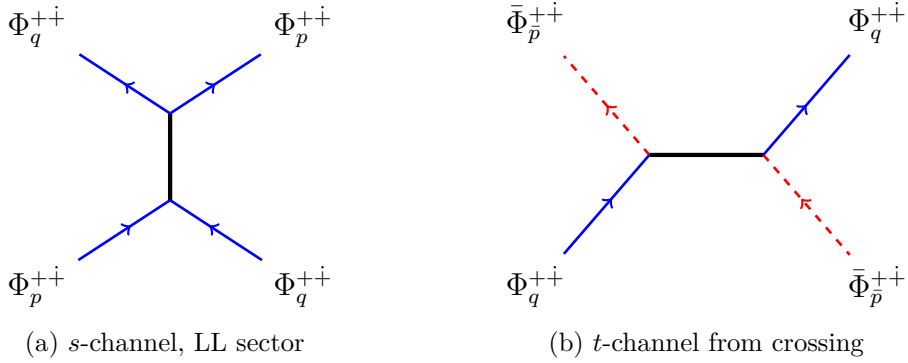
\begin{figure}%
  \centering%
  \subfloat[$s$-channel, LL sector]{%
    \label{fig:s-channel-LL}
    \begin{tikzpicture}[
      thick,
      level/.style={level distance=1.15cm},
      level 2/.style={sibling distance=3.5cm},
      ]
      \coordinate
      child[grow=up]{
        edge from parent [draw=black,line width=1.6]
        child {
          node{$\Phi^{+\dot{+}}_p$}
          edge from parent [particle]
        }
        child {
          node{$\Phi^{+\dot{+}}_q$}
          edge from parent [particle]
        }
        node [above=3pt]{}
      }
      child[grow=down, level distance=0pt] {
        child {
          node{$\Phi^{+\dot{+}}_p$}
          edge from parent [antiparticle]
        }
        child { 
          node{$\Phi^{+\dot{+}}_q$}
          edge from parent [antiparticle]               
        }
      };
    \end{tikzpicture}%
  }%
  \hspace{2cm}%
  \subfloat[$t$-channel from crossing]{%
    \label{fig:t-channel-LR}%
    \begin{tikzpicture}[
      thick,
      level/.style={level distance=1.5cm},
      level 2/.style={sibling distance=3.5cm},
      ]
      \coordinate
      child[grow=left]{
        edge from parent [draw=black,line width=1.6]
        child {
          node{$\bar{\Phi}^{+\dot{+}}_{\bar{p}}$}
          edge from parent [particlecross]
        }
        child {
          node{${\Phi}^{+\dot{+}}_{q}$}
          edge from parent [antiparticle]
        }
        node [above=3pt]{}
      }
      child[grow=right, level distance=0pt] {
        child {
          node{$\bar{\Phi}^{+\dot{+}}_{\bar{p}}$}
          edge from parent [antiparticlecross]
        }
        child { 
          node{${\Phi}^{+\dot{+}}_{q}$}
          edge from parent [particle]               
        }
      };
    \end{tikzpicture}}%
  \caption{%
    On the left two particles in the same sector form an~$\algSU(2)$ bound state in the $s$-channel. Applying the crossing transformation to~$\Phi_p^{+\dot{+}}$ yields the $t$-channel diagram on the right, where on particle has unphysical momentum~$\bar{p}$ (red dashed lines). Particles are labeled as in~\cite{Borsato:2013qpa}.
  }%
  \label{fig:LLpoles}%
\end{figure}

Scattering processes involving formation or exchange of bound states give rise to single poles in the S-matrix for physical values of the spectral parameters. Let us consider the $s$-channel diagram in figure~\subref*{fig:s-channel-LL}. The process involves two fundamental particles from the same sector, \eg two left-movers, in the physical region $|x_i|>1$, $i=p,q$, which form an on-shell boundstate and then split up again. 
Similarly to the case of the $\algSU(2)$ sector in $\AdS_5$~\cite{Dorey:2007xn}, this should lead to a pole in the corresponding S-matrix element at $x_p^+=x_q^-$.
The relevant element is\footnote{Here we write the S-matrix elements in the ``spin-chain frame''. Accounting for the frame factors does not modify the pole structure~\cite{Borsato:2013qpa}.}
\begin{equation}
\mathcal{A}_{pq}=\bra{\Phi^{+\dot{+}}_q \Phi^{+\dot{+}}_p }\Smat \ket{\Phi^{+\dot{+}}_p \Phi^{+\dot{+}}_q}  = \frac{x_p^- - x_q^+}{x_p^+ - x_q^-} \frac{1-\frac{1}{x_p^- x_q^+}}{1-\frac{1}{x_p^+ x_q^-}} \sigma_{pq}^{-2}.
\end{equation}
As discussed in appendix~\ref{sing-dressing} the dressing factor is regular at $x_p^+=x_q^-$, so that $\mathcal{A}_{pq}$ has a simple pole there.

This $s$-channel process is related through crossing symmetry to the exchange of a bound state in the $t$-channel, depicted in figure~\subref*{fig:t-channel-LR}. There the particle of momentum~$p$ has been crossed so that $x_{\bar{p}}^\pm=1/x_p^\pm$ are not in the physical region. Since the two processes are related by crossing symmetry, the poles in the $s$-channel automatically fix the singularities in the $t$-channel. In fact crossing symmetry implies~\cite{Borsato:2013qpa}
\begin{equation}
\mathcal{A}_{pq} \widetilde{\mathcal{A}}_{\bar{p}q}=1, \qquad \text{  where  } \widetilde{\mathcal{A}}_{pq}=\bra{\Phi^{+\dot{+}}_q \bar{\Phi}^{+\dot{+}}_p }\Smat \ket{\bar{\Phi}^{+\dot{+}}_p \Phi^{+\dot{+}}_q}\,,
\end{equation}
so that a pole of $\mathcal{A}_{pq}$ corresponds to a pole of ${\widetilde{\mathcal{A}}_{\bar{p}q}}^{-1}$. We can check this explicitly by considering 
\begin{equation}
\widetilde{\mathcal{A}}_{\bar{p}q} = \frac{1-\frac{1}{x_{\bar{p}}^- x_q^+}}{1-\frac{1}{x_{\bar{p}}^- x_q^-}} \,  \frac{1-\frac{1}{x_{\bar{p}}^+ x_q^-}}{1-\frac{1}{x_{\bar{p}}^+ x_q^+}} \tilde{\sigma}_{\bar{p}q}^{-2}.
\end{equation}
Since $\tilde{\sigma}$ is regular when continued inside the unit circle (see appendix \ref{sing-dressing}), $\widetilde{\mathcal{A}}_{\bar{p}q}$ has a zero at $x_{\bar{p}}^+=1/x_q^-$, as expected.

If we consider S-matrix elements involving one left- and one right-moving particle we expect no poles, since there are no corresponding bound states. Therefore a process such as the one depicted in figure~\subref*{fig:s-channel-LR} should not happen. Indeed, the S-matrix element
\begin{figure}
  \centering%
  \subfloat[RL $s$-channel (forbidden)]{%
    \label{fig:s-channel-LR}%
    \begin{tikzpicture}[
      thick,
      level/.style={level distance=1.15cm},
      level 2/.style={sibling distance=3.5cm},
      ]
      \coordinate
      child[grow=up]{
        edge from parent [draw=black,line width=1.6]
        child {
          node{$\bar{\Phi}^{-\dot{-}}_p$}
          edge from parent [particle]
        }
        child {
          node{$\Phi^{+\dot{+}}_q$}
          edge from parent [particle]
        }
        node [above=3pt]{}
      }
      child[grow=down, level distance=0pt] {
        child {
          node{$\bar{\Phi}^{-\dot{-}}_p$}
          edge from parent [antiparticle]
        }
        child { 
          node{$\Phi^{+\dot{+}}_q$}
          edge from parent [antiparticle]               
        }
      };
    \end{tikzpicture}%
  }%
  \hspace{2cm}%
  \subfloat[crossed LL $t$-channel (forbidden)]{%
    \begin{tikzpicture}[
      thick,
      level/.style={level distance=1.5cm},
      level 2/.style={sibling distance=3.5cm},
      ]
      \coordinate
      child[grow=left]{
        edge from parent [draw=black,line width=1.6]
        child {
          node{${\Phi}^{-\dot{-}}_{\bar{p}}$}
          edge from parent [particlecross]
        }
        child {
          node{$\Phi^{+\dot{+}}_{q}$}
          edge from parent [antiparticle]
        }
        node [above=3pt]{}
      }
      child[grow=right, level distance=0pt] {
        child {
          node{${\Phi}^{-\dot{-}}_{\bar{p}}$}
          edge from parent [antiparticlecross]
        }
        child { 
          node{$\Phi^{+\dot{+}}_{q}$}
          edge from parent [particle]               
        }
      };
    \end{tikzpicture}
    \label{fig:t-channel-LL}}
  \caption{%
    On the left the would-be Landau diagram for one left- and one right-moving particle is depicted. This process should be absent. Similarly, the crossed process on the right should be absent, and the corresponding S-matrix element have no pole. 
  }%
  \label{fig:LRpoles}%
\end{figure}
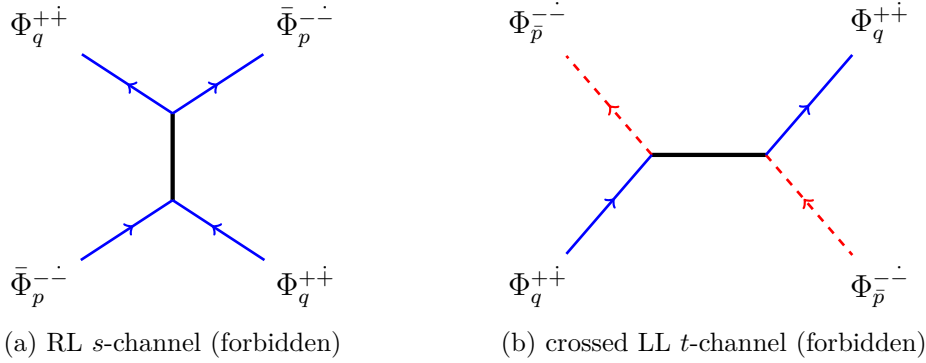%
\begin{equation}
\widetilde{\mathcal{B}}_{pq}=\bra{\Phi^{+\dot{+}}_q \bar{\Phi}^{-\dot{-}}_p }\Smat \ket{\bar{\Phi}^{-\dot{-}}_p \Phi^{+\dot{+}}_q}  =  \frac{1-\frac{1}{x_p^- x_q^+}}{1-\frac{1}{x_p^+ x_q^-}} \,  \frac{1-\frac{1}{x_p^+ x_q^+}}{1-\frac{1}{x_p^- x_q^-}} \tilde{\sigma}_{pq}^{-2},
\end{equation}
is regular in the physical region, and in particular has no pole at $x_p^+=x_q^-$.
It is an interesting check that the same holds in the crossed channel, whose exchange diagram would be as in figure~\subref*{fig:t-channel-LL}, should it exist. 
Again crossing symmetry relates the two processes by
\begin{equation}
\widetilde{\mathcal{B}}_{pq} \mathcal{B}_{\bar{p}q}=1, \qquad \text{where}\quad \mathcal{B}_{pq}=\bra{\Phi^{+\dot{+}}_q \Phi^{-\dot{-}}_p }\Smat \ket{\Phi^{-\dot{-}}_p \Phi^{+\dot{+}}_q}\,,
\end{equation}
which implies the first crossing equation in~\eqref{eq:crossing12}.
Since $\widetilde{\mathcal{B}}_{pq}$ has no singularity at $x_p^+=x_q^-$ we expect $\mathcal{B}_{\bar{p}q}$ to have no singularity at $x_{\bar{p}}^+=1/x_q^-$. Explicitly we have 
\begin{equation}
\mathcal{B}_{\bar{p}q} = \frac{(x_{\bar{p}}^- - x_q^-)^2}{(x_{\bar{p}}^- - x_q^+)(x_{\bar{p}}^+ - x_q^-)} \, \frac{1-\frac{1}{x_{\bar{p}}^- x_q^+}}{1-\frac{1}{x_{\bar{p}}^+ x_q^-}}  \sigma_{\bar{p}q}^{-2}.
\end{equation}
The rational terms have a pole at $x_{\bar{p}}^+=1/x_q^-$, but  once the dressing factor is continued to the crossed region as in~\eqref{eq:chizero}, this is canceled by a zero of $\sigma^{-2}$, so that the result is non-singular.

The solutions to the crossing equations that we found can be modified by multiplying them by some ``CDD factors''. Similarly to the $\AdS_5$ case~\cite{Volin:2009uv}, we expect them to be meromorphic functions of the spectral parameters that solve the homogeneous crossing equations
\begin{equation}
\sigma^{\CDD}_{pq}\,\widetilde{\sigma}^{\CDD}_{\bar{p}q}=1\,,\qquad
\sigma^{\CDD}_{\bar{p}q}\,\widetilde{\sigma}^{\CDD}_{pq}=1\,.
\end{equation}
Such factors would introduce pairs of zeros and poles in the two phases, for instance by letting
\begin{equation}
\chi^{\CDD}_{pq}=\frac{i}{2}\log\frac{(x-y)^{c_1}}{(1-xy)^{c_2}}\,,\qquad
\widetilde{\chi}^{\CDD}_{pq}=\frac{i}{2}\log\frac{(x-y)^{c_2}}{(1-xy)^{c_1}}\,,
\end{equation}
with $c_1,\,c_2$ integer constants. However, since the pole structure of the dressing factors and of the resulting S-matrix agrees with the one we expect from the bound state content of the theory we can consistently set
\begin{equation}
\sigma^{\CDD}_{pq}=1\,,\qquad\widetilde{\sigma}^{\CDD}_{pq}=1\,,
\end{equation}
so that the full dressing phases are given by~\eqref{eq:solution}, up to non-trivial solutions of the homogeneous crossing equations with no poles in the physical region.

\section{Expansions of the dressing factors}\label{sec:expansions}
In section~\ref{sec:crossing-solution} we solved the $\AdS_3$ crossing equations~(\ref{eq:crossing1}) in terms of~(\ref{eq:solution}).
In what follows we give the strong- and weak-coupling expansions of these all-loop phases.

\subsection{Strong-coupling expansion}\label{sec:expansion-strong}
In this subsection we compute the strong coupling expansion of the dressing phases in order to compare with the perturbative string theory calculations of~\cite{Beccaria:2012kb}.
The dressing phases have an expansion in terms of local conserved charges $q_r(p_k)$~\cite{Arutyunov:2004vx}
\begin{equation}\label{eq:theta-charges}
  \theta(p_1,p_2) = \sum_{r=1}^{\infty} \sum^\infty_{\substack{s>r \\ r+s=\text{odd}}} c_{r,s}(h) \left[ q_r(p_1)q_s(p_2)-q_r(p_2)q_s(p_1) \right]\,,
\end{equation}
where $c_{r,s}(h)$ are functions of the coupling constant $h$ with expansion
\begin{equation}
c_{r,s}(h)=h c_{r,s}^{(0)}+ c_{r,s}^{(1)}+ c_{r,s}^{(2)}h^{-1}+\dots
\label{eq:cexp}
\end{equation} 
and are antisymmetric in $r,s$. The phase $\widetilde{\theta}(p_1,p_2)$ has a similar expansion where the coefficients will be denoted $\tilde{c}_{r,s}(h)$. The expression above is similar to the corresponding one in $\AdS_5$, but unlike that case, we will need to include the $r=1$ terms. This new feature was first noted in~\cite{Beccaria:2012kb}. For $r\geq 2$ the conserved charges are given by
\begin{equation}
q_r(p_k) = 
\newq_r(x_k^+)-\newq_r(x_k^-)= 
\frac{i}{r-1} \left[ \frac{1}{(x_k^+)^{r-1}} - \frac{1}{(x_k^-)^{r-1}}\right]\,, 
\qquad \newq_r(x)\equiv \frac{i}{r-1} \frac{1}{x^{r-1}},
\end{equation}
where we introduced the function $\newq_r(x_k)$ for later convenience.
For $r=1$ the charge is just the momentum
\begin{equation}
q_1(p_k) = \newq_1(x_k^+)-\newq_1(x_k^-)=-i \log{\left(\frac{x_k^+}{x_k^-}\right)}, \qquad \newq_1(x)\equiv i \log\left(\frac{1}{x}\right)\,.
\end{equation}
Expressing $\theta(p_1,p_2)$ in terms of $\chi$ (\cf equation~\eqref{eq:thchi}), we obtain the expansion
\begin{equation}
  \chi(x,y) = \sum_{r=1}^{\infty} \sum^\infty_{\substack{s>r \\ r+s=\text{odd}}} c_{r,s}(h) \left[ \newq_r(x)\newq_s(y) - \newq_r(y)\newq_s(x) \right].
  \label{eq:chiexp}
\end{equation}
with a corresponding expression for $\widetilde{\chi}$.
The coefficients $c_{r,s}$ and $\tilde{c}_{r,s}$ can be obtained by expanding the integrands through which $\chi$ and $\tilde{\chi}$ are defined at large $h$ and at large $x$ and $y$ and then performing the integrals. The expansions for $\chi^{\BES}$ and $\chi^{\HL}$ are well known in the literature, and in particular we have
\begin{equation}
\chi^{\HL}(x,y)=\frac{2}{\pi}\sum_{r=2}^\infty\sum^\infty_{\substack{s>r \\ r+s=\text{odd}}}\frac{(r-1)(s-1)}{(r-s)(r+s-2)}\left[ \newq_r(x)\newq_s(y) - \newq_r(y)\newq_s(x) \right]
\,.
\label{eqchihlcrs}
\end{equation}
The expansion for $\chi^-$ (\cf equation~\eqref{eq:chi-}) is
\begin{equation}
  \begin{aligned}
    \chi^-(x,y) &= -\frac{1}{\pi}\sum_{r=2}^\infty\sum^\infty_{\substack{s>r \\ r+s=\text{odd}}}\frac{(r-1)^2+(s-1)^2}{(r-s)(r+s-2)}\left[ \newq_r(x)\newq_s(y) - \newq_r(y)\newq_s(x) \right]
    \\ &\phantom{{}={}}
    +\frac{1}{2\pi}\sum^\infty_{\substack{s>1 \\ s=\text{even}}}\left[ \newq_1(x)\newq_s(y) - \newq_1(y)\newq_s(x) \right]\,.
    \label{eqchimcrs}
  \end{aligned}
\end{equation}
Expanding~(\ref{eq:solution}) at large $h$ we find
\begin{equation}
\begin{aligned}
\chi(x,y) &= h\, \chi^{\AFS}(x,y) + \frac{1}{2} ( \chi^{\HL}(x,y) + \chi^{-}(x,y) ) + \mathcal{O}\big(\frac{1}{h}\big)\,, \\ 
\widetilde{\chi}(x,y) &= h\, \chi^{\AFS}(x,y) + \frac{1}{2} ( \chi^{\HL}(x,y) - \chi^{-}(x,y) ) + \mathcal{O}\big(\frac{1}{h}\big)\,,
\end{aligned}
\end{equation}
where we have extracted the $h$-scaling of each phase. At leading order both phases reduce to the AFS one, as it was found in the perturbative string expressions obtained in~\cite{Beccaria:2012kb}, so that
\begin{equation}
c^{(0)}_{\BLMMT\, r,s}={\bar c}^{(0)}_{\BLMMT\, r,s}=c^{(0)}_{r,s}=\tilde{c}^{(0)}_{r,s}=\delta_{r+1,s}\,.
\end{equation}

At HL-order all three terms on the r.h.s of equation~(\ref{eq:solution}) contribute and we find
\begin{equation}
\begin{aligned}
c_{r,s}^{(1)}&= +\frac{1}{2\pi} \frac{1-(-1)^{s+r}}{2}\left[\frac{s-r}{s+r-2}-\frac{1}{2}\big(\delta_{r,1}-\delta_{1,s}\big)\right]\,,\\
\tilde{c}_{r,s}^{(1)}&= -\frac{1}{2\pi} \frac{1-(-1)^{s+r}}{2}\left[\frac{s+r-2}{s-r}-\frac{1}{2}\big(\delta_{r,1}-\delta_{1,s}\big)\right]\,,
\end{aligned}
\end{equation}
for $s>r>0$. Comparing to the semiclassical results~\cite{Beccaria:2012kb} we find, for $r>1$
\begin{equation}
c^{(1)}_{\BLMMT\, r,s}= 4 \pi c^{(1)}_{r,s}\,,\qquad\qquad
{\bar c}^{(1)}_{\BLMMT\, r,s}=4 \pi \tilde{c}^{(1)}_{r,s}\,.
\end{equation}
The factors of $4 \pi$ are there since~\cite{Beccaria:2012kb} expand in $\sqrt{\lambda}$ and we expand in $h=\frac{\sqrt{\lambda}}{4\pi}$ (see equation~\eqref{eq:cexp}). In summary, the $r>1$ coefficients at HL-order give the same contribution to the dressing phases as those found in~\cite{Beccaria:2012kb}. On the other hand, the $r=1$ coefficients, which come exclusively from the expansion of~$\chi^-(x,y)$ are
\begin{eqnarray}
c^{(1)}_{\BLMMT\, 1,s}=
8 \pi c^{(1)}_{1,s}
\,,\qquad\qquad
{\bar c}^{(1)}_{\BLMMT\, 1,s}=
8\pi \tilde{c}^{(1)}_{1,s}\,.
\label{cr1disc}
\end{eqnarray}
Taking into account the factor of $4\pi$ discussed above, we conclude that the $r=1$ coefficients of~\cite{Beccaria:2012kb} give \emph{twice} the contribution found here. 

One possible origin for this discrepancy could have to do with the antisymmetrisation procedure used in~\cite{Beccaria:2012kb}. Unlike $r>1$ terms, the $r=1$ contributions to the BA dressing phases can be simplified using the momentum conservation condition. As such, this part of the phase need not be explicitly antisymmetric. In the next subsection we discuss another possible likely source of this discrepancy. 

Finally, the higher order coefficients~$c^{(n)}_{r,s}=\tilde{c}^{(n)}_{r,s}$ with $n>1$ are exactly the same as in the expansion of the BES phase.

\subsection{Semiclassical and near flat space limits}\label{sec:semiclassical-and-NFS}
In order to compare with perturbative results, it is convenient to write explicit expressions for our phases in the semiclassical limit, \ie, when
\begin{equation}
x^\pm =x\pm\frac{i}{2h}\frac{x^2}{x^2-1}+O\left(\frac{1}{h^3}\right)\,.
\end{equation}
Such an expansion for the BES phase is well known: the leading order~$O(1/h)$ is given by the AFS phase~\eqref{eq:AFS-xpxm}, which in our normalization reads
\begin{equation}
\theta^{\AFS}(x,y)=\frac{1}{h}\frac{x-y}{(x^2-1)(x y-1)(y^2-1)}+O\left(\frac{1}{h^3}\right)\,,
\end{equation}
whereas the next-to-leading-order is given by the HL phase which can be found by expanding~\eqref{eq:DHM-HL} under the integral. Doing so also for~\eqref{eq:chi-}, we get to the expressions
\begin{equation}
\begin{aligned}
\theta(x,y)=\theta^{\AFS}(x,y)+\frac{1}{4\pi h^2}\frac{x^2}{x^2-1}&\frac{y^2}{y^2-1}\big[\frac{(x+y)^2(1-\frac{1}{xy})}{(x^2-1)(x-y)(y^2-1)}\\
&\ +\frac{2}{(x-y)^2}\log\big(\frac{x+1}{x-1}\frac{y-1}{y+1}\big)\big]+O\left(\frac{1}{h^3}\right),\\
\widetilde{\theta}(x,y)=\theta^{\AFS}(x,y)+\frac{1}{4\pi h^2}\frac{x^2}{x^2-1}&\frac{y^2}{y^2-1}\big[\frac{(xy+1)^2(\frac{1}{x}-\frac{1}{y})}{(x^2-1)(xy-1)(y^2-1)}\\
&\ + \frac{2}{(xy-1)^2}\log\big(\frac{x+1}{x-1}\frac{y-1}{y+1}\big)\big]+O\left(\frac{1}{h^3}\right).
\end{aligned}
\label{eq:FGsol}
\end{equation}
As it was expected from the discussion in~\cite{Borsato:2013qpa}, the rational part of these expression differs from the one found in~\cite{Beccaria:2012kb}. The logarithmic part agrees\footnote{
One should keep into account a factor of $-2$ coming from the different definition of the phases in~\cite{Beccaria:2012kb}.
} with what conjectured in~\cite{Beccaria:2012kb}, also in agreement with recent results found by unitarity techniques~\cite{Engelund:2013fja,Bianchi:2013nra}. The discrepancy in the rational part may come from the fact that the Bethe ansatz assumed in~\cite{Beccaria:2012kb} differs from the one of~\cite{Borsato:2013qpa} by terms of the form
\begin{equation}
\frac{1-\frac{1}{x^+y^-}}{1-\frac{1}{x^-y^+}}\frac{1-\frac{1}{x^+y^+}}{1-\frac{1}{x^-y^-}}\qquad\text{or}
\qquad\frac{1-\frac{1}{x^+y^-}}{1-\frac{1}{x^-y^+}}\frac{1-\frac{1}{x^-y^-}}{1-\frac{1}{x^+y^+}}\,.
\end{equation}
The former term in each product is antisymmetric, and can just be absorbed by a redefinition of the phase $\widetilde{\sigma}$. This is not true for the latter term which is \emph{symmetric}. This will contribute to the Bethe ansatz in the finite gap limit as
\begin{equation}
\frac{1-\frac{1}{x^+y^+}}{1-\frac{1}{x^-y^-}}=\exp\left({\frac{i}{h}\frac{x+y}{(x^2-1)(y^2-1)}}\right)+O\left(\frac{1}{h^3}\right)\,.
\end{equation}
such a contribution has presumably to be taken into account before antisymmetrisation and regularization procedure performed in~\cite{Beccaria:2012kb} and may nontrivially affect it.

Let us also evaluate the dressing factors in the near-flat-space limit~\cite{Maldacena:2006rv}
\begin{equation}
\begin{aligned}
\theta(p_-,q_-)=\frac{p_- q_- (p_--q_-)}{16 h (p_-+q_-)}+\frac{p_-^2 q_-^2 \left(p_-^2+2 p_- q_- \log\frac{q_-}{p_-}-q_-^2\right)}{256 \pi  h^2
   (p_--q_-)^2}+O\left(\frac{1}{h^3}\right)\,,\\
\widetilde{\theta}(p_-,q_-)=\frac{p_- q_- (p_--q_-)}{16 h (p_-+q_-)}-\frac{p_-^2 q_-^2 \left(p_-^2-2 p_- q_- \log\frac{q_-}{p_-}-q_-^2\right)}{256 \pi  h^2
   (p_-+q_-)^2}+O\left(\frac{1}{h^3}\right)\,.
\end{aligned}
\label{eq:NFSsol}
\end{equation}
Using these results one can check that the dressing phases are compatible with the near-flat-space results of~\cite{Sundin:2013ypa}. 

\subsection{Weak-coupling expansion}
In this subsection we compute the weak-coupling expansion of the dressing phases. The results for $\sigma^{\BES}$ are well known from $\AdS_5/\CFT_4$. The leading-order contribution to the dressing phase starts at $\order(h^6)$~\cite{Beisert:2006ez}, and comes from the $r=2$, $s=3$ terms in the expansion of $\chi^{\BES}$.\footnote{See equation~\eqref{eq:chiexp}, and recall that for the BES phase there is no $r=1$ term.} 

The $\AdS_3$ dressing phases~\eqref{eq:solution} contain extra terms besides the BES phase. The coefficients $c_{r,s}$ and $\tilde{c}_{r,s}$ that come from these extra contributions are all order $h^0$ (see equation~\eqref{eqchimcrs} and~\eqref{eqchihlcrs}). The coupling constant dependence comes only from the charges $q_r$ and $q_s$ in equation~\eqref{eq:chiexp}. In fact, the leading contribution comes from the $r=1$ and $s=2$ term
\begin{equation}
\theta(p,q)= 4 c^{(1)}_{1,2} \left(p \sin^2\frac{q}{2} - q \sin^2\frac{p}{2} \right) h + \order(h^3) \,,
\end{equation}
with a similar expression holding for~$\widetilde{\theta}(p,q)$. Note that the~$O(h^2)$ terms vanish.
The above result shows that the $r=1$ terms, which are novel to $\AdS_3$, contribute at order $h$ to the BA,\footnote{%
  Notice that, despite being linear in one of the momenta, such terms cannot be re-absorbed into a shift of the Bethe Ansatz length, since~$c^{(1)}_{1,2}=-\tilde{c}^{(1)}_{1,2}$ so that they appear with opposite sign in~$\theta$ and~$\widetilde{\theta}$.%
} %
and so should modify the energy of states in the weakly-coupled spin-chain at order $h^3$. 
Notice that \emph{a priori} we do not know how $h(\lambda)$ behaves at weak coupling.\footnote{%
  Recall, for example, that in $\AdS_5/\CFT_4$ $h\sim\sqrt{\lambda}$ while in $\AdS_4/\CFT_3$ $h\sim\lambda$.%
} %
This prevents us from determining whether the $h^1$ contribution to $\theta(p,q)$ in the equation above comes with an integral power of $\lambda$ as one would expect in a weakly coupled planar limit. Nevertheless the above expansion is a new feature of the $AdS_3$ spin-chain, which places it in a different category to the spin chains investigated in~\cite{Beisert:2005wv}. This is not surprising, since in contrast to~\cite{Beisert:2005wv}, the $AdS_3$ spin-chain consists of left-moving and  right-moving 
sectors.\footnote{%
  Something similar happens in the study of general alternating spin-chains~\cite{Bargheer:2009xy}, where novel operators that do not exist for the homogeneous spin-chains investigated in~\cite{Beisert:2005wv} modify the structure of the dressing phase found in~\cite{Beisert:2005wv}.%
} %
Spin-chains with a left- and right-moving copy of a symmetry group will have a larger family of operators that can act on them than the homogeneous chains of~\cite{Beisert:2005wv}, and it would be interesting to extend the analysis of~\cite{Beisert:2005wv} to this case, in order to better understand the role of the $r=1$ terms in the dressing phase.

\section{Conclusions}

We have determined the non-perturbative dressing phases of the $\AdS_3/\CFT_2$ integrable system associated to Type IIB string theory on $\AdS_3\times\Sphere^3\times \Torus^4$ with R-R flux. This was done by solving the crossing relations of~\cite{Borsato:2013qpa}. Our solution differs from the BES dressing phase that enters $\AdS_5/\CFT_4$ and $\AdS_4/\CFT_3$ integrable systems. The two phases we have found are different from one another as is expected from the crossing equations. We have investigated the spectrum of bound states of the system and show that it is consistent with the full non-perturbative S-matrix. The details of this matching depend crucially on the analytic properties of the dressing phases. As such, this represents a strong consistency check of our solution

We have performed an expansion of the dressing phases at strong coupling. At the leading order both phases reduce to the AFS-phase, in agreement with perturbative world-sheet calculations~\cite{Rughoonauth:2012qd,Beccaria:2012kb,Sundin:2013ypa}. At the next-to-leading order our phases differ from one another and only their sum is the same as the HL-phase. We have compared our expressions at this order with the results of~\cite{Beccaria:2012kb,Sundin:2013ypa} and found almost complete agreement. In section~\ref{sec:expansion-strong} and~\ref{sec:semiclassical-and-NFS} we discussed the likely origins of the discrepancy.

In order to further check whether our solutions correspond to the string theory phases, it is necessary to test them against stringent perturbative calculations, beyond the HL-order.
In addition, studying their analytical properties in the string and mirror regions may give further insights on the validity of our proposal.\footnote{We thank Sergey Frolov for his remarks on this point.}
It would also be very interesting to investigate the double poles/zeros of our phases and compare them to relevant Landau diagrams as was done in $\AdS_5/\CFT_4$ in~\cite{Dorey:2007xn}. Another important direction would be to build on~\cite{Sax:2012jv} in order to understand how massless modes should be incorporated into the integrable S-matrix. There is by now significant evidence that integrable spin-chains play an important role in the context of $\AdS_3/\CFT_2$. Finding the origin of such spin-chains in the $\CFT_2$ remains an outstanding challenge.

\section*{Acknowledgments} 

We would like to thank Gleb Arutyunov, Matteo Beccaria, Niklas Beisert, Gustav Delius, Nick Dorey, Sergey Frolov, Fedor Levkovich-Maslyuk, Tomasz \L ukowski, Guido Macorini, Andrea Prinsloo, Arkady Tseytlin, Dymitro Volin and Kostya Zarembo for interesting discussions, and Gleb Arutyunov, Arkady Tseytlin and Kostya Zarembo for their comments on the manuscript.
R.B., O.O.S.\@ and A.S.\@ acknowledge support by the Netherlands Organization for Scientific Research (NWO) under the VICI grant 680-47-602; their work is also part of the ERC Advanced grant research programme No. 246974, ``Supersymmetry: a window to non-perturbative physics''. 
B.S.\@ acknowledges funding support from from an EPSRC Advanced
Fellowship and an STFC Consolidated Grant ``Theoretical Physics at City University''  ST/J00037X/1. He would also like to thank the CERN Theory division for hospitality during the initial stages of this project.
A.T.\@ thanks EPSRC for funding under the First Grant project EP/K014412/1 ``Exotic quantum groups, Lie superalgebras and integrable systems''.

\appendix
\section{Useful formulae and identities}
\label{app:identities}
In this appendix we present the proofs of some identities we used in the main body of the paper. In particular, we provide the proof that the phase $\chi^-(x,y)$ solves the crossing equation~\eqref{eq:cr-ratio}. This can be easily adapted to check that the Hern\'andez-L\'opez phase as defined in~\eqref{eq:intu-intd-HL} solves the ``odd'' crossing equation~\eqref{eq:cr-HL} with the choice of path depicted in figure~\ref{fig:toruspaths}.

\subsection{Solving equation~(\ref{eq:cr-ratio})}
\label{app:solving}
\noindent Let us define the following integral
\begin{equation}\label{eq:Phi-}
\begin{aligned}
\Phi^-(x,y) &=\ointc \, \frac{dw}{8\pi} \frac{1}{x-w} \log{\left[ (y-w)\left(1-\frac{1}{yw}\right)\right]} \, \text{sign}((w-1/w)/i) \ - x \leftrightarrow y \\
&=\left( \inturl - \intdlr \right)\frac{dw}{8\pi} \frac{1}{x-w} \log{\left[ (y-w)\left(1-\frac{1}{yw}\right)\right]} \ - x \leftrightarrow y,
\end{aligned}
\end{equation}
which is reminiscent of~(\ref{eq:intu-intd-HL}). This function satisfies a property which will be crucial in what follows, that is\footnote{The proof of this is presented in appendix~\ref{app:identities-proof}.}
\begin{equation}
\label{eq:phi-id}
\Phi^-(x,y)-\Phi^-(1/x,y)= 0.
\end{equation}
Furthermore, when $|y|>1$ is fixed and $|x|\geq1$ approaches the unit circle, $\Phi^-(x,y)$ has a jump discontinuity. As discussed in appendix~\ref{app:identities-disc}, the value of the discontinuity depends on whether $x$ approaches the unit circle form below the real line, in which case
\begin{equation}
\Phi^-(e^{i\varphi+\epsilon},y)=\Phi^-(e^{i\varphi-\epsilon},y)+\delta_{\uparrow}(e^{i\varphi},y)+O(\epsilon),\qquad  \epsilon>0,\quad -\pi<\varphi<0\,,
\end{equation}
with\footnote{More precisely, the following relation holds up to an arbitrary function of $y$ only, and in an appropriate branch of the logarithm, see~\ref{app:identities-disc}. Such a functions plays no role in the crossing equation.}
\begin{equation}
\label{eq:deltaminus}
\delta_{\uparrow}(x,y)=-\frac{i}{2}\,\log\left[(y-x)\left(1-\frac{1}{xy}\right)\right]\,,
\end{equation}
or from above, where 
\begin{equation}
\Phi^-(e^{i\varphi+\epsilon},y)=\Phi^-(e^{i\varphi-\epsilon},y)+\delta_{\downarrow}(e^{i\varphi},y)+O(\epsilon),\qquad  \epsilon>0,\quad 0<\varphi<\pi\,,
\end{equation}
with $\delta_{\downarrow}(x,y)=-\delta_{\uparrow}(x,y)$.

These ingredients are all we need to construct a solution of~\eqref{eq:cr-ratio}. In the physical region we define
\begin{equation}
\chi^-(x,y)\equiv\Phi^-(x,y)\,\qquad |x|,|y|>1\,.
\end{equation}
In order to continue this function to the crossed region, it is important to recall our choice of cuts of figure~\ref{fig:toruspaths}: both~$x^+(z)$ and~$x^-(z)$ will cross the unit circle \emph{below the real line}. Therefore, we define
\begin{equation}
\chi^-(x,y)\equiv\Phi^-(x,y)+\delta_{\uparrow}(x,y)\,\qquad |x|<1,\quad |y|>1\,,
\end{equation}
which is continuous across the lower half circle by construction. Using~\eqref{eq:phi-id} we have 
\begin{equation}
\chi^-(x,y)-\chi^-(1/x,y)=-\delta_{\uparrow}(1/x,y)=-\delta_{\uparrow}(x,y)\,,\qquad |x|,|y|>1\,.
\end{equation}
Rewriting the left-hand-side of~\eqref{eq:cr-ratio} in terms of $\chi(x,y)$ gives finally
\begin{equation}
\label{eq:crossdifffinal}
\frac{{\ratiosigma(z_1,z_2)}^2}{{\ratiosigma(z_1+\omega_2,z_2)}^2}=\frac{e^{-2i (\delta_{\uparrow}(x^+,y^+)+ \delta_{\uparrow}(x^-,y^-))}}{e^{-2i (\delta_{\uparrow}(x^+,y^-)+\delta_{\uparrow}(x^-,y^+))}}\,,
\end{equation}
 which coincides with~\eqref{eq:cr-ratio-expl}.

\subsection{Identity for \texorpdfstring{$\Phi^-(x,y)-\Phi^-(1/x,y)$}{Phi(x,y) - Phi(1,x,y)}}
\label{app:identities-proof}

In this subsection we prove equation~\eqref{eq:phi-id}. Define
\begin{equation}
\label{eq:phisplit}
F(x,y)=\Fup(x,y)-\Fdw(x,y)=\inturl f(w,x,y)dw - \intdlr f(w,x,y)dw\,,
\end{equation}
where $\Fup,\,\Fdw$ corresponds to the first and second integral, respectively, and
\begin{equation}
f(w,x,y)=\frac{1}{8\pi} \frac{1}{x-w} \log{\left[ (y-w)\left(1-\frac{1}{yw}\right)\right]}\,,
\end{equation}
so that $\Phi^-(x,y)=F(x,y)-F(y,x)$. Since $f(w,x,y)-f(w,x,1/y)=0$, we see that
\begin{equation}
\Fup(x,y)-\Fup(x,1/y)=0\,,\qquad \mbox{and}\qquad
\Fdw(x,y)-\Fdw(x,1/y)=0\,.
\label{eq:proof1overy}
\end{equation}
A change of integration variable, $u=1/w$, can be used to derive the following identity
\begin{align}
  \Fdw(1/x,y)&=\intdlr \frac{dw}{8\pi} \frac{1}{1/x-w} \log{\left[ (y-w)\left(1-\frac{1}{yw}\right)\right]} \nonumber \\
  &=\inturl \frac{du}{8\pi\,u^2} \frac{1}{1/x-1/u} \log{\left[ (y-u)\left(1-\frac{1}{yu}\right)\right]} \label{eq:proof1overx} \\
  &=-\inturl \frac{du}{8\pi} \frac{1}{x-u} \log{\left[ (y-u)\left(1-\frac{1}{yu}\right)\right]}
  -\inturl \frac{du}{8\pi} \frac{1}{u} \log{\left[ (y-u)\left(1-\frac{1}{yu}\right)\right]} \nonumber \\
  &=-\Fup(x,y)-\phi^-(y)\,, \nonumber
\end{align}
for arbitrary $x$ with $|x|\neq1$. Sending $x\to1/x$ in the above equation gives
\begin{equation}
\Fup(1/x,y)=-\Fdw(x,y)-\phi^-(y)\,.
\label{eq:proof1overxeq2}
\end{equation} 
Combining equations~(\ref{eq:proof1overy}),~(\ref{eq:proof1overx}) 
and~(\ref{eq:proof1overxeq2}) and writing out $\Phi$ in terms of $\Fup$ and $\Fdw$ one may check that~\eqref{eq:phi-id} holds.

\subsection{Discontinuities of \texorpdfstring{$\Phi^-(x,y)$}{Phi(x,y)} at \texorpdfstring{$|x|=1$}{|x|=1}}
\label{app:identities-disc}
Let us split $\Phi^-(x,y)$ in terms of $\Fup,\,\Fdw$ as in~\eqref{eq:phisplit}, and focus on the discontinuities of~$\Fdw(x,y)$. The discontinuity in $x$ follows immediately from Cauchy's theorem, and is given by
\begin{equation}
\Fdw(e^{i\varphi+\epsilon},y)=\Fdw(e^{i\varphi-\epsilon},y)+d^{(x)}_{\uparrow}(e^{i\varphi},y)+O(\epsilon),\qquad  \epsilon>0,\quad -\pi<\varphi<0\,,
\end{equation}
crossing the unit circle from below, with
\begin{equation}
d^{(x)}_{\uparrow}(x,y)=\frac{i}{4}\,\log\left[(y-x)\left(1-\frac{1}{xy}\right)\right]\,.
\end{equation}
Note that $\Fdw(x,y)$ is continuous in $x$ across the upper half-circle.

To find the discontinuity in $y$, we can consider $\partial_y\Fdw(x,y)$ for $|x|,|y|>1$; bringing the derivative under the integral gets rid of the logarithm. The resulting function has a discontinuity on the lower half circle:
\begin{equation}
\partial_y\Fdw(x,e^{i\varphi+\epsilon})=\partial_y\Fdw(x,e^{i\varphi-\epsilon})+d^{(y')}_{\uparrow}(x,e^{i\varphi})+O(\epsilon),\qquad  \epsilon>0,\quad -\pi<\varphi<0\,.
\end{equation}
The appropriate primitive of $d^{(y')}_{\uparrow}(x,y)$ in $y$ gives the discontinuity of $\Fdw(x,y)$ from below. Such a primitive is
\begin{equation}
d^{(y)}_{\uparrow}(x,y)=-\frac{i}{4}\,\log\left(x-y\right)+\phi_\uparrow(x)\,,
\end{equation}
where $\phi_\uparrow(x)$ is arbitrary function. Furthermore, there is also a discontinuity on the upper half circle:
\begin{equation}
\partial_y\Fdw(x,e^{i\varphi+\epsilon})=\partial_y\Fdw(x,e^{i\varphi-\epsilon})+d^{(y')}_{\downarrow}(x,e^{i\varphi})+O(\epsilon),\qquad  \epsilon>0,\quad 0<\varphi<\pi\,,
\end{equation}
whose primitive is
\begin{equation}
d^{(y)}_{\downarrow}(x,y)=\frac{i}{4}\,\left(\log\left(1-xy\right)-\log x -\log y\right)+\phi_\downarrow(x)\,.
\end{equation}

It is easy to repeat this analysis for $\Fup(x,y)$, where we find essentially the same results up to exchanging the upper and lower circles. Putting everything together proves~\eqref{eq:deltaminus} up to such an arbitrary function of $x$, which however would drop out of the crossing equation~\eqref{eq:crossdifffinal}, canceling among the contributions of the four $\chi$'s. Since the crossing relations~\eqref{eq:crossdifffinal} are written in exponential form, the logarithmic branch cuts of the discontinuities play no role. 

If we instead had considered the discontinuities of $\Phi^-(x,y)$ when crossing the unit circle in the \emph{upper} half-plane we would have found an extra minus sign upon crossing.

\subsection{Singularities of the dressing phases}\label{sing-dressing}
Let us investigate the singularities of the dressing phases~$\theta(x,y)$ and~$\widetilde{\theta}(x,y)$. They are defined in terms of $\chi(x,y),\,\widetilde{\chi}(x,y)$ by~\eqref{eq:solution}. Since the analytic properties of the BES phase are well known~\cite{Dorey:2007xn,Arutyunov:2009kf}, we will focus on the semisum and semidifference of the HL phase with $\chi^-(x,y)$. We are interested in logarithmic singularities as~$x$ and~$y$ take particular positions with respect to each other. To find them, let us consider the integrals
\begin{equation}
\label{eq:PsiPM}
\Psi^\pm(x,y)=\frac{1}{2}\big(-\Phi^{\text{HL}}(x,y)\pm\Phi^-(x,y)\big)\,,
\end{equation}
where $\Phi^{\text{HL}}(x,y)$ is the integral defining the HL phase in the physical region,
\begin{equation}
\Phi^{\text{HL}}(x,y)=\left( \inturl - \intdlr \right)\frac{dw}{4\pi} \frac{1}{x-w} \left(\log{(y-w)}-\log{\left(y-1/w\right)}\right),
\end{equation}
and $\Phi^-(x,y)$ is defined in~\eqref{eq:Phi-}. Singularities may arise only at~$y=x$ or~$y=1/x$. However, we can evaluate explicitly
\begin{equation}
\Psi^\pm(x,y)\big|_{y=x}=0\,,\qquad\Psi^\pm(x,y)\big|_{y=1/x}=\frac{1}{4\pi}\big(4\,\text{Li}_2(x)-\text{Li}_2(x^{2})\big)\,,
\end{equation}
with $|y|>1$, so that no singularity arises from the integral representation. Since this coincides with the phase in the physical region, we can conclude that there is no discontinuity for the phases at $x=y$ when both variables are in the physical region. 

When $y=1/x$ one of the variables, \eg, $x$, must be inside the unit circle. Therefore, as explained in appendix~\ref{app:solving}, we must continue $\Psi^\pm(x,y)$ in $x$ through the lower half-circle. In order to do this we need to find the discontinuity of $\Psi^\pm(x,y)$ there. In appendix~\ref{app:identities-disc} we worked out the discontinuity of~$\Phi^-(x,y)$ to be as in~\eqref{eq:deltaminus}, \ie,
\begin{equation}
\delta_{\uparrow}^-(x,y)=-\frac{i}{2}\,\log\left[(y-x)\left(1-\frac{1}{xy}\right)\right]\,.
\end{equation}
Using Cauchy's theorem we find that $\Phi^{\text{HL}}(x,y)$ satisfies
\begin{equation}
\Phi^{\text{HL}}(e^{i\varphi+\epsilon},y)=\Phi^{\text{HL}}(e^{i\varphi-\epsilon},y)+\delta_{\uparrow}^{\text{HL}}(e^{i\varphi},y)+O(\epsilon),\qquad  \epsilon>0,\quad -\pi<\varphi<0\,,
\end{equation}
with
\begin{equation}
\delta_{\uparrow}^{\text{HL}}(x,y)=-\frac{i}{2}\,\log\left[\frac{y-x}{y-1/x}\right]\,.
\end{equation}
Using this to analytically continue~\eqref{eq:PsiPM} we have that when $|x|<1$ and $|y|>1$, there is no singularity in $\widetilde{\chi}(x,y)$ at $y=1/x$. However $\chi(x,y)$ has a logarithmic singularity such that
\begin{equation}
\label{eq:chizero}
e^{2i \chi(x,y)}\approx \big(y-\frac{1}{x}\big)\,,\qquad\text{when}\quad y\approx 1/x\,.
\end{equation}

\section{Choice of analytic continuation}
\label{sec:an-cont}
Using the S-matrix derived in \cite{Borsato:2013qpa}, one can write down two sets of crossing equations, which turn out to be incompatible. This problem is not specific to our case, but also appears, \eg, for $\AdS_5$. These two possibilities are related to the fact that charge conjugation can be implemented either with $C$ or $C \, \Sigma$. In the first case the first entry has to be analytically continued by $z+\omega_2$, while in the second case by $z-\omega_2$. The opposite is true for the second entry.
 
We write these crossing equations for $\AdS_3 \times \Sphere^3 \times \Torus^4$ in the following table. In the first column we write the crossing equations explicilty in terms of $\sigma, \tilde{\sigma}$ (the functions $g,\tilde{g}$ are defined in~\eqref{eq:crossingF}) and in the second column we write them in matrix form.
\begin{center}
  \begin{tabular}{ @{\rule{2pt}{0pt}} c @{\rule{2pt}{0pt}} >{\small} c @{\rule{4pt}{0pt}} c @{\rule{0pt}{0pt}}}
    \toprule \\[-4pt]
    \multirow{2}{*}{\scriptsize \rule{0pt}{14pt}(I)} &
    $\sigma(z_1+\omega_2,z_2)^2 \tilde{\sigma}(z_1,z_2)^2 = g(x_1^\pm,x_2^\pm)$ & \\[4pt]
    & $\sigma(z_1,z_2)^2 \tilde{\sigma}(z_1+\omega_2,z_2)^2 = \tilde{g}(x_1^\pm,x_2^\pm)$ & 
    \multirow{-2}{*}{\small \rule[-8pt]{0pt}{8pt} $C^{-1}\otimes \matId\cdot\mat{R}^{\text{t}_1}(z_1+\omega_2,z_2)\cdot C\otimes\matId={\mat{R}(z_1,z_2)}^{-1}$}
    \\[12pt]%
    \multirow{2}{*}{\scriptsize \rule{0pt}{18pt}(II)} &
    $\sigma(z_1,z_2-\omega_2)^2 \tilde{\sigma}(z_1,z_2)^2 = \tilde{g}^{-1} \Bigl(\frac{1}{x_2^\pm},x_1^\pm\Bigr)$ & \\[8pt]%
    & $\sigma(z_1,z_2)^2 \tilde{\sigma}(z_1,z_2-\omega_2)^2 = g^{-1}\Bigl(\frac{1}{x_2^\pm},x_1^\pm\Bigr)$ & 
    \multirow{-2}{*}{\small \rule[-14pt]{0pt}{14pt} $\matId\otimes C^{-1}\cdot \mat{R}^{\text{t}_2}(z_1,z_2-\omega_2)\cdot\matId\otimes C={\mat{R}(z_1,z_2)}^{-1}$} 
    \\[12pt]%
    \midrule\\[-8pt]
    \multirow{2}{*}{\scriptsize \rule{0pt}{14pt}(III)} &
    $\sigma(z_1-\omega_2,z_2)^2 \tilde{\sigma}(z_1,z_2)^2 = g(x_1^\pm,x_2^\pm) \vphantom{\Bigl(} $& \\[4pt]%
    & $\sigma(z_1,z_2)^2 \tilde{\sigma}(z_1-\omega_2,z_2)^2 =\tilde{g}(x_1^\pm,x_2^\pm) \vphantom{\Bigl(}$ & 
    \multirow{-2}{*}{\small \rule[-10pt]{0pt}{10pt} $\Sigma C^{-1}\otimes \matId\cdot\mat{R}^{\text{t}_1}(z_1-\omega_2,z_2)\cdot C\Sigma\otimes\matId={\mat{R}(z_1,z_2)}^{-1}$}
    \\[12pt]%
    \multirow{2}{*}{\scriptsize \rule{0pt}{18pt}(IV)} &
    $\sigma(z_1,z_2+\omega_2)^2 \tilde{\sigma}(z_1,z_2)^2 = \tilde{g}^{-1}\Bigl(\frac{1}{x_2^\pm},x_1^\pm\Bigr)$ & \\[8pt]%
    & $\sigma(z_1,z_2)^2 \tilde{\sigma}(z_1,z_2+\omega_2)^2 = g^{-1}\Bigl(\frac{1}{x_2^\pm},x_1^\pm\Bigr)$ & 
    \multirow{-2}{*}{\small \rule[-14pt]{0pt}{14pt} $\matId\otimes \Sigma C^{-1}\cdot \mat{R}^{\text{t}_2}(z_1,z_2+\omega_2)\cdot\matId\otimes C  \Sigma={\mat{R}(z_1,z_2)}^{-1}$}
    \\[12pt]
    \bottomrule
  \end{tabular}
\end{center}
Looking at the first column, it is clear that rows (I) and (III) are incomplatible because the r.h.s is the same but the analytic continuation is performed in opposite directions. The same is true for rows (II) and (IV).
Rows (I) and (II) are instead related by using antisymmetry of the factors, as are (III) and (IV).
 
In order to fix the convention for analytic continuation we compare our results with the perturbative results of~\cite{Sundin:2013ypa}.
We consider the matrix elements $\mathcal{A}_{pq}, \widetilde{\mathcal{A}}_{pq}, \mathcal{B}_{pq}, \widetilde{\mathcal{B}}_{pq}$ defined in section~\ref{sec:S-matrix-poles} and we write the crossing equations in the form
\begin{equation}
\mathcal{A}_{pq} \, \widetilde{\mathcal{A}}_{\bar{p}q}=1, \qquad \widetilde{\mathcal{B}}_{pq} \, \mathcal{B}_{\bar{p}q}  =1.
\end{equation}
In order for the equations to be satisfied up to one-loop order in the NFS limit, we use that $\bar{p}_-=-p_-$ and we note that we need to choose the branch of the log in such a way that $\log(-p_-) = \log(p_-)-i \pi$, which also implies the rule $\sqrt{-p_-} = -i \, \sqrt{p_-}$.
We can relate these choices of branches to the choice of the sign of the shift on the torus by considering the analytic continuation of
\begin{equation}\label{eq:eta-xpm}
\eta(p) = \left( \frac{x_p^+}{x_p^-} \right)^{1/4} \left( \frac{i h}{2} \left(x_p^- - x_p^+\right) \right)^{1/2}=\frac{\dn\frac{z}{2}\,\big(\cn\frac{z}{2}+i\,\sn\frac{z}{2}\,\dn\frac{z}{2}\big)}{1+16h^2\,\sn^{4}\frac{z}{2}}\,,
\end{equation}
which is given by~\cite{Arutyunov:2009ga}
\begin{equation}\label{eq:eta-an-cont}
\eta(z \pm \omega_2) = \pm \frac{i}{x^+(z)} \eta(z).
\end{equation}
By using~\eqref{eq:eta-xpm} we find that in the NFS limit $\eta(p_-)=1/2 \sqrt{p_-}$. We can now compare our choice of the log branch with the crossing transformation, finding
\begin{equation}
\eta(-p_-)=\frac{1}{2} \sqrt{-p_-}= -\frac{i}{2} \, \sqrt{p_-}=\eta(z + \omega_2)\big|_{\text{NFS~limit}}\,.
\end{equation}
This allows us to conclude that crossing holds with a shift by $+\omega_2$ in the first variable. The crossing equations in row (I) in the table are the ones that are solved in the main text. Consistency with this choice allows us to conlcude that the crossing equations written in~\cite{Borsato:2012ud,Borsato:2012ss} should follow the same convention of shifting by $-\omega_2$ in the second variable. This corrects footnote 10 in~\cite{Borsato:2012ud} and footnote 3 in~\cite{Borsato:2012ss}.

A posteriori we can check in the same way that the solutions we found in the finite gap~\eqref{eq:FGsol} and near flat space limit~\eqref{eq:NFSsol} do indeed solve the crossing equations in those limits when shifting the first variable by $z\to z+\omega_2$ and resolving the branch of the log accordingly.

\bibliographystyle{nb}
\bibliography{crossing}

\end{document}